\documentclass[nopreprintline,letterpaper,11pt,numbers,sort&compress]{elsarticle}

\usepackage{mathrsfs}
\usepackage{ifpdf}
\usepackage{amsmath}
\usepackage{amsthm}
\usepackage{psfrag}
\usepackage{overpic}
\usepackage{rotating}
\usepackage{graphicx,amssymb,lineno}
\usepackage{subfigure}
\usepackage{array}

\usepackage[text={6.5in,8.5in},centering]{geometry} 
\newcolumntype{L}[1]{>{\raggedright\let\newline\\\arraybackslash\hspace{0pt}}m{#1}}
\newcolumntype{R}[1]{>{\raggedleft\let\newline\\\arraybackslash\hspace{0pt}}m{#1}}
\newcommand{\ba}{\begin{array} }
\newcommand{\ea}{\end{array} }
\newcommand{\bae}{\begin{eqnarray}}
\newcommand{\eae}{\end{eqnarray}}
\newcommand{\bea}{\begin{eqnarray*}}
\newcommand{\eea}{\end{eqnarray*}}
\newcommand{\be}{\begin{equation}}
\newcommand{\ee}{\end{equation}}

\usepackage{amsfonts}
\usepackage{amssymb}
\usepackage{amsthm}
 \usepackage{ae} 
\usepackage[T1]{fontenc}
\usepackage[ansinew]{inputenc}
\usepackage{amsmath}
\usepackage{eucal}

\usepackage[titletoc,title]{appendix} 
\usepackage[english]{babel}
\usepackage{soul}

\usepackage{color}

\usepackage{lscape}
\usepackage{graphicx}
\usepackage{epstopdf}
\ifpdf
\usepackage[%
  pdftitle={Instructions for use of the document class
    elsart},%
  pdfauthor={},%
  pdfsubject={The preprint document class elsart},%
  pdfkeywords={instructions for use, elsart, document class},%
  pdfstartview=FitH,%
  bookmarks=true,%
  bookmarksopen=true,%
  breaklinks=true,%
  colorlinks=true,%
  linkcolor=blue,anchorcolor=blue,%
  citecolor=blue,filecolor=blue,%
  menucolor=blue,pagecolor=blue,%
  urlcolor=blue]{hyperref}
\else
\usepackage[%
  breaklinks=true,%
  colorlinks=true,%
  linkcolor=blue,anchorcolor=blue,%
  citecolor=blue,filecolor=blue,%
  menucolor=blue,pagecolor=blue,%
  urlcolor=blue]{hyperref}
\fi

\usepackage{secdot}

\AtBeginDocument{}

\newcommand{\blue}[1]{\textcolor{black}{#1}}

\newcommand{\cyan}[1]{\textcolor{black}{#1}}

\newtheorem{thm}{Theorem}[section]
\newtheorem{lem}{Lemma}[section]
\newtheorem{prop}{Proposition}[section]

\usepackage{lipsum}
\makeatletter
\def\ps@pprintTitle{%
 \let\@oddhead\@empty
 \let\@evenhead\@empty
 \def\@oddfoot{}%
 \let\@evenfoot\@oddfoot}
\makeatother

\begin{document}
\begin{frontmatter}

\title{Nutritional Regulation Influencing Colony Dynamics and Task Allocations in Social Insect Colonies}

\author[1]{Feng Rao}\ead{raofeng2002@163.com}
\address[1]{School of Physical and Mathematical Sciences, Nanjing Tech University, Nanjing, Jiangsu 211816, China}
\author[2]{Marisabel Rodriguez Messan}\ead{marisabel@asu.edu}
\address[2]{Department of Ecology and Evolutionary Biology, Brown University, Providence RI 02912 USA}
\author[3]{Angelica Marquez}\ead{amarquez45@miners.utep.edu}
\address[3]{College of Engineering, University of Texas at El Paso, El Paso, Texas, USA.}
\author[4]{Nathan Smith}\ead{nesmith6@asu.edu}
\address[4]{School of Life Sciences, Arizona State University, Tempe, AZ 85287, USA}
\author[5]{Yun Kang}\ead{yun.kang@asu.edu}
\address[5]{College of Integrative Sciences and Arts, Arizona State University, Mesa, AZ 85212, USA}

\begin{abstract}

In this paper, we use an adaptive modeling framework to model and study how nutritional status (measured by the protein to carbohydrate ratio) may regulate population dynamics and foraging task allocation of social insect colonies.  Mathematical analysis of our model shows that both investment to brood rearing and brood nutrition are important for colony survival and dynamics. When division of labor and/or nutrition are in an intermediate value range, the model undergoes a backward bifurcation and creates multiple attractors due to bistability. This bistability implies that there is a threshold population size required for colony survival. When the investment in brood is large enough or nutritional requirements are less strict the colony tends to survive, otherwise the colony faces collapse. Our model suggests that the needs of colony survival are shaped by the brood survival probability, which requires good nutritional status. As a consequence, better nutritional status can lead to a better survival rate of larvae, and thus a larger worker population. \\
\end{abstract}


\end{frontmatter}

\section{Introduction}\label{S-1}

In social insect colonies such as ants, bees and wasps, all members of the colony work collectively to ensure colony survival. Colonies act as a single common organism capable of making decisions and forming complex behavioral connections between its members~\cite{kang2015ecological,KangY2016BMB}. They exhibit a decentralized system  with a sophisticated division of labor resulting from interactions among members of the colony and the environment~\cite{Camazine2001book,beshers2001models,KangY2016BMB}.  In addition to the reproductive division of labor between the queen and the workers, workers also have a division of labor between foragers which leave the nest to search for food and non-foragers that carry out tasks within the nest \cite{daugherty2011nutrition}. In social insect societies, foraging responsibilities are assigned to a subset of adult colony members \cite{cook2010colony}. Internal and external factors happening at both the individual and colony level shape the foragers' decision to bring back a certain type of food \cite{cook2010colony}. \\

Currently, there are few studies that have focused on the outcome of nutrient regulation in social insects at the colony level \cite{cook2010colony} (but see \cite{dussutour2008carbohydrate,dussutour2009communal,toth2005nutritional}). Many of these studies lack focus on the overall outcome of colony population dynamics, and how nutrient regulation among foragers affects the number of reared brood, mortality of adult workers, and in general colony survival. In this study, we focus on the mechanisms that regulate foraging behavior of eusocial workers and the outcomes of these mechanisms on colony performance, including but not limited to the number of brood raised and worker mortality. The collection of food resources by an indivdiual forager is based not only on the colony's current nutritional status, but also on the worker's physical caste, age, and prior experience \cite{Traniello1989ARE,pohl2016colony}. The nutritional needs of the colony are shaped by the differing needs of larvae and workers in the colony \cite{cook2010colony,dussutour2008description,pohl2016colony}. For instance, the growth of larvae relies heavily on protein, while worker ants require primarily carbohydrates as a source of energy \cite{dussutour2008description,markin1970food,wilson1957quantitative,cassill1998emergent,cassill1999regulation,Dussutour2012PRSB}. Many studies have shown that the ratio of protein to carbohydrates in the diet of a range of insect species is crucial for performance \cite{cook2010colony,dussutour2008description,raubenheimer1999integrating,simpson2004optimal,lee2008lifespan}, though, in general, carbohydrates are often more attractive to foragers than protein \cite{pohl2016colony,dussutour2008carbohydrate}. However, the protein required for growth may be in greater demand when a queen is laying eggs \cite{pohl2016colony}. \\

In order for social insect foragers to compensate for potential nutrient restrictions in the food available to the colony \cite{pohl2016colony,seeley1989social,cassill1999information,clark2011behavioral}, foragers adjust their collection in favor of food sources containing limiting nutrients \cite{pohl2016colony,dussutour2008carbohydrate,clark2011behavioral}. This guarantees that the colony meets its longer term objectives and thus promotes colony growth and reproduction \cite{pohl2016colony}. According to Dussutour \textit{et al.} \cite{dussutour2008carbohydrate}, within a colony, workers recruit nestmates for food collection at different rates depending upon food type \cite{cassill1999regulation,portha2002self}, food concentration, and hunger level \cite{mailleux2006starvation,dussutour2008carbohydrate}. At the individual level, when workers are starved recruitment will be stronger to carbohydrate-rich food sources than to sources high in protein \cite{dussutour2008carbohydrate}. At a collective level, deployment of foragers to carbohydrate-rich or highly proteinaceous material increases in the presence of larvae, resulting in an increase in the collection of carbohydrates and protein \cite{dussutour2008carbohydrate,portha2002self,brian1972population}.\\

There are several empirical studies that have studied how a colony is affected by the availability of required nutrients for colony growth and reproduction, and how workers regulate collection of these nutrients to meet individual and collective demands~\cite{dussutour2008carbohydrate,dussutour2009communal,cook2010colony,Dussutour2012PRSB,Lihoreau2015,Pohl2016}. However, currently there are no mathematical models to our knowledge that have attempted to study these mechanisms dynamically. The main goal of this paper is to propose and study an adaptive modeling framework to further understand how nutritional status may regulate population dynamics and foraging task allocation of social insect colonies. The proposed model contains three compartments that allow us to analyze and measure the impacts of nutritional status that can benefit colony growth and survival. Our model assumes that  (1) nutritional status is measured by the protein to carbohydrate ratio, which reflects the ratio of workers foraging for protein to workers foraging for carbohydrates; (2) brood are able to survive if the protein to carbohydrate ratio falls into a certain range; and (3) the colony recruits workers to forage for protein or carbohydrate in order to maximize the brood survival rate. In addition, our proposed model includes division of labor implicitly. Also, by considering the basic mechanisms affecting colony growth such as cooperative effort for reproductive division of labor, successful brood maturation/survival, and recruitment of workers to collect different nutrients based on specific colony nutritional demands, our model could help us understand how other life history factors affect the performance (number of brood raised and mortality of workers) of the colony.  \\

The rest of this article is organized as follows: In section \ref{S-2}, we describe the detailed derivation of our proposed model. In section \ref{mathanalysis} we provide the mathematical analysis of our model including lemmas, propositions, and theorems, the proofs of which can be found in section \ref{proofs}. In section \ref{simulation}, we provide numerical simulations illustrating the equilibrium dynamics of the model to further obtain biological insights of some life history parameters of the colony. Lastly, the conclusion of this paper is found in section \ref{S-Dis}.\\

\section{Derivation of the mathematical model}\label{S-2}
 
Let $L(t)$ represent the brood population; $A_p(t)$ be the portion of foragers collecting proteinaceous material, called the protein forager; $A_c(t)$ be the portion of foragers collecting carbohydrates, called the carbohydrate forager. The total forager population is denoted as $A(t)=A_p(t)+A_c(t)$. The following ecological assumptions determine the population of $L,\,A_c$ and $A_p$:

\begin{enumerate}
\item \textbf{Brood population $L(t)$:}
The brood population $L$ increases with the average egg-laying rate of the queen(s) given by $\gamma$, which is discounted by two factors:
\begin{enumerate}
\item The survival rate function of eggs is determined by the cooperative efforts of workers $A$ in the colony. We adopt the modeling approach from Kang \textit{et al.}~\cite{kang2011mathematical,Messan2017MMNP}, where the cooperative efforts that lead to the eggs' survival is measured by a Holling type-III function $\frac{aA^2}{b+aA^2}$, where $b$ is a half-saturation constant and $a$ is the portion of the division of labor invested towards the successful development of the larvae.
\item The survival rate of larvae to workers is determined by the available nutrients in the colony which is reflected through the protein to carbohydrate ratio of worker collectors $S_L\left(\frac{A_p}{A_c}\right)$. Examples of $S_L$ could be $S_L\left(\frac{A_p}{A_c}\right)=-\alpha_1\Big|\frac{A_p}{A_c}-\theta_m\Big|+\alpha_2(\theta_c-\theta_m)$ with $\alpha_i\in(0,1)\>i=1,2$ as a scaling factor of nutrient collection, $\theta_m$ representing the optimal nutrient ratio, and $\theta_c$ representing the maximal nutrient ratio that brood can survive (see Figures~\ref{function}\subref{function-general-1}), or general functions such as the normal biological performance curve (see Figure \ref{function}\subref{function-general-2}). Notice that $S_L\leq 1$ can be negative, thus we define $S_{L_{\max}}=\max\{0,S_L\}$ such that ${S_{L_{\max}}}\in [0,1]$ is a survival probability. 
\end{enumerate}
The brood population decreases by a maturation rate $\beta L$, which describes the rate at which brood matures into the adult class $A$.  Thus, we have following equation:

 \[L'=\gamma \cdot \underbrace{S_{L_{\max}}}_{\text{nutrient effects}}  \cdot\underbrace{\frac{aA^2}{b+aA^2}}_{\text{adult worker efforts}}-\underbrace{\beta L}_{\text{maturation rate}}.\]
\vspace{12pt}

When $\frac{A_p}{A_c}$ is less than $\theta_m$, the brood survival rate increases, and decreases when $\frac{A_p}{A_c}$ is greater than $\theta_m$. This phenomenon has been supported by the work of \cite{cook2010colony, lee2008lifespan,dussutour2009communal,kay2006ant}, in which it is explained that worker survivability decreases as a probable side effect of an excess ingestion of proteins and of carbohydrate limitation. Figure~\ref{function}\subref{function-general-1} shows a general case of $S_L\left(\frac{A_p}{A_c}\right)=-\alpha_1\Big|\frac{A_p}{A_c}-\theta_m\Big|+\alpha_2(\theta_c-\theta_m)$ with different $\alpha_1$ and $\alpha_2$. 
The partial derivative of $S_L=S_L\left(\frac{A_p}{A_c}\right)$ reveals constant rates, showing a linear relation between the brood survival rate and nutritional status. In this study, we assume that when the nutrition level hits or exceeds the critical value $\theta_c$, i.e., $\frac{A_p}{A_c}\geq \theta_c$, the nutrient becomes toxic such that no brood can survive. Lastly, Figure~\ref{function}\subref{function-general-2} shows how the survival rate grows with respect to the collection of nutrients until it reaches $\theta_m$. 

\begin{figure}[!ht]
\begin{center}
\subfigure[A general case of $S_L\left(\frac{A_p}{A_c}\right)$] 
{\includegraphics[scale=0.36]{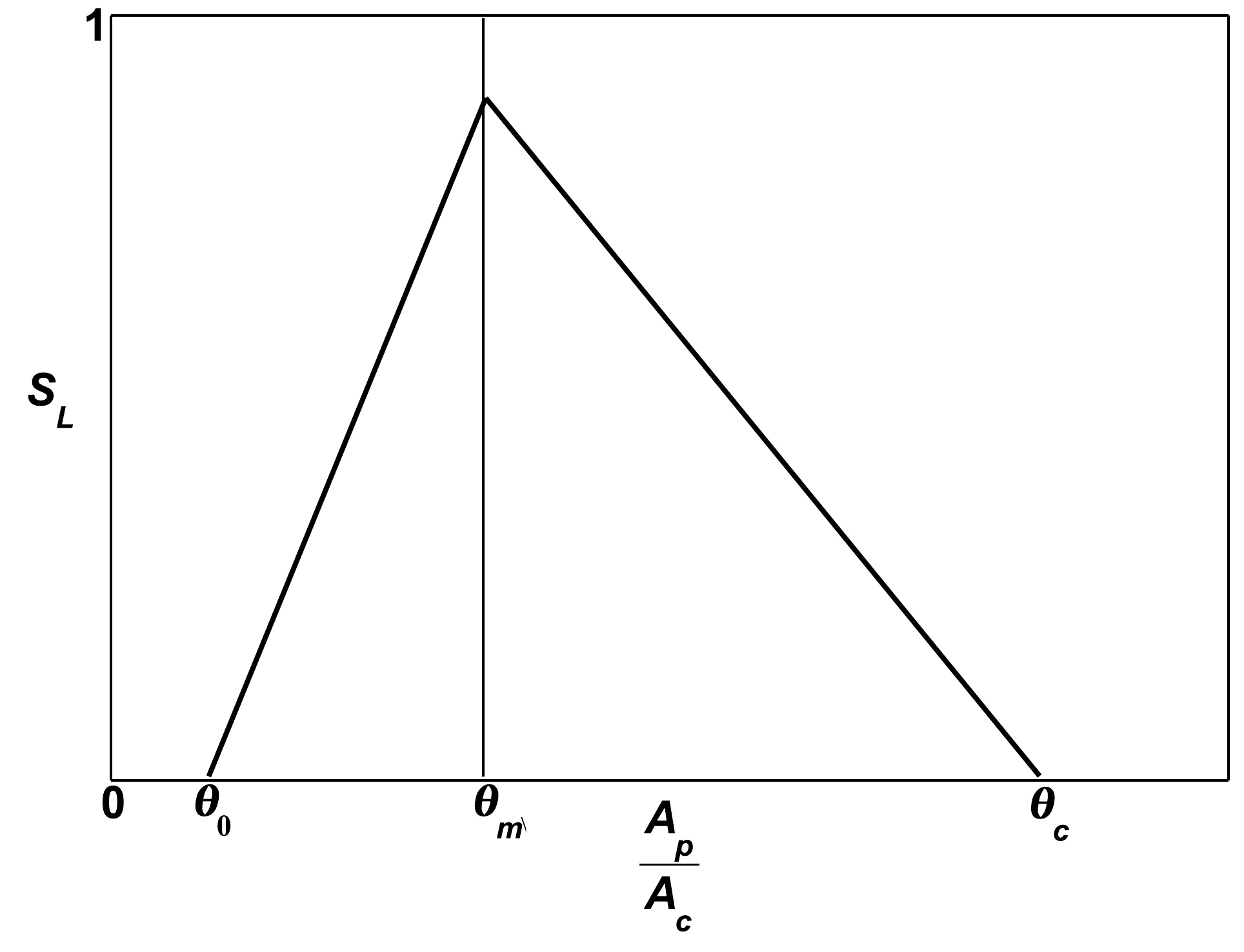}\label{function-general-1}}\hspace{0.7cm}
\subfigure[Normal biological performance curve $S_L\left(\frac{A_p}{A_c}\right)$] 
{\includegraphics[scale=0.36]{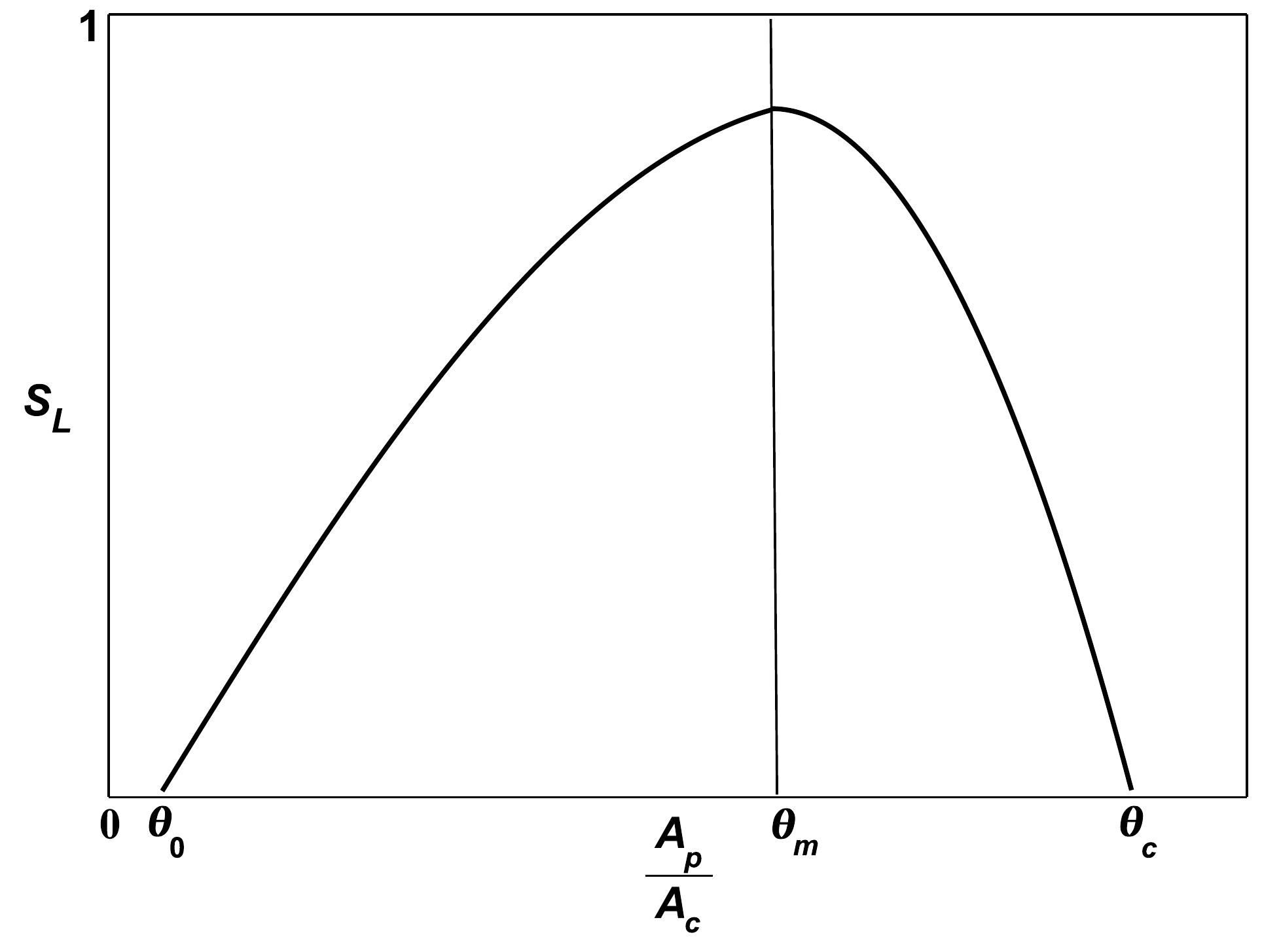}\label{function-general-2}}
\end{center}
\vspace{-20pt}
\caption{Examples of a survival rate function: (a) A general case of $S_L\left(\frac{A_p}{A_c}\right)=-\alpha_1\Big|\frac{A_p}{A_c}-\theta_m\Big|+\alpha_2(\theta_c-\theta_m)$ with different $\alpha_1=0.3$ and $\alpha_2=0.15$; (b) The normal biological performance curve $S_L\left(\frac{A_p}{A_c}\right)$.}
\label{function}
\end{figure}

In general, we expect the protein to {carbohydrate} ratio $\frac{A_p}{A_c}$ of the colony to fall in a certain range  in order for the colony to survive and grow, say, $\frac{A_p}{A_c}\in [\theta_0,\theta_m]$, where $\theta_0$ is the minimum nutrient ratio necessary for brood survival. This is supported by \cite{cook2010colony,dussutour2009communal}. Thus it is reasonable to assume that $S_{L_{\max}}\left(\frac{A_p}{A_c}\right)$ has the following simple form:
\begin{equation}\label{SLg}
S_{L_{\max}}\left(\frac{A_p}{A_c}\right)=\left\{\begin{array}{ll}
0 &\mbox{ when } 0\leq \frac{A_p}{A_c}<\theta_0\\
\alpha_1 \left(\frac{A_p}{A_c}-\theta_0\right) &  \mbox{ when\, } \theta_0\leq \frac{A_p}{A_c}<\theta_m\\
-\alpha_2 \left(\frac{A_p}{A_c}-\theta_c\right)  & \mbox{ when\, } \theta_m\leq \frac{A_p}{A_c}<\theta_c\\
0& \mbox{ when } \theta_c\leq \frac{A_p}{A_c}
\end{array}\right.
\end{equation}subject to ${\alpha_i} \in(0,1), i=1,2$ and $\alpha_1 \left(\theta_m-\theta_0\right) +\alpha_2 \left(\theta_m-\theta_c\right)=0 \mbox{ and } 0<\alpha_1 \left(\theta_m-\theta_0\right)\leq 1.$

In particular, we have 
$$S_L=\alpha_1\left(\frac{A_p}{A_c}-\theta_0\right)  \mbox{ when\, }  \frac{A_p}{A_c}< \theta_m;\,\,S_L=-\alpha_2\left(\frac{A_p}{A_c}-\theta_c\right)  \mbox{ when\, }  \frac{A_p}{A_c}\geq \theta_m.$$
Then we have 
\begin{equation}\label{Sl-der}
\begin{array}{ll}
\frac{\partial S_L}{\partial\frac{A_p}{A_c}}=\alpha_1;\, &\frac{\partial S_L}{\partial\frac{A_c}{A_p}}=-\alpha_1 (\frac{A_p}{A_c})^2<0\mbox{ \,when\, }  \frac{A_p}{A_c}<\theta_m\\
\frac{\partial S_L}{\partial\frac{A_p}{A_c}}=-\alpha_2;\,&\frac{\partial S_L}{\partial\frac{A_c}{A_p}}=\alpha_2 (\frac{A_p}{A_c})^2>0 \mbox{ \,when\, } \theta_m\leq \frac{A_p}{A_c}<\theta_c
\end{array}
\end{equation}

In the symmetric case, i.e., $\alpha_1=\alpha_2=\alpha$, then we have 
$$S_L\left(\frac{A_p}{A_c}\right)=-\alpha\Big|\frac{A_p}{A_c}-\theta_m\Big|+\alpha(\theta_c-\theta_m) \mbox{ when } \frac{A_p}{A_c}\in [\theta_0, \theta_c]$$ with $\alpha\in(0,1)$, $\theta_0=2\theta_m-\theta_c$ and $\theta_c\in [\theta_m, 2\theta_m+\frac{1}{\alpha}]$. Thus, $S_L=\alpha (\frac{A_p}{A_c}+\theta_c-2\theta_m)$ if $2\theta_m-\theta_c\leq\frac{A_p}{A_c}<\theta_m$ and $S_L=\alpha (\theta_c-\frac{A_p}{A_c})$ if $\theta_m\leq \frac{A_p}{A_c}<\theta_c$. In addition, we have the following:

\begin{equation}\label{Sl-der-s}
\begin{array}{ll}
\frac{\partial S_L}{\partial\frac{A_p}{A_c}}=\alpha;\, &\frac{\partial S_L}{\partial\frac{A_c}{A_p}}=-\alpha (\frac{A_p}{A_c})^2\mbox{ \,when\, } 2\theta_m-\theta_c\leq \frac{A_p}{A_c}<\theta_m,\\[.3cm]
\frac{\partial S_L}{\partial\frac{A_p}{A_c}}=-\alpha;\,&\frac{\partial S_L}{\partial\frac{A_c}{A_p}}=\alpha (\frac{A_p}{A_c})^2 \mbox{ \,when\, } \theta_m\leq \frac{A_p}{A_c}<\theta_c.
\end{array}
\end{equation}

The special case of the symmetric scenario above is $S_L(0)=0$, i.e., $\theta_c=2\theta_m$. In this case, we have $S_L\left(\frac{A_p}{A_c}\right)=-\alpha\Big|\frac{A_p}{A_c}-\theta_m\Big|+\alpha \theta_m$ with $\theta_m\in (0, \frac{1}{\alpha})$ in our proposed model \eqref{M1}. In the following section, we will provide mathematical analysis of the general case of $S_{L_{\max}}\left(\frac{A_p}{A_c}\right)$ shown in \eqref{SLg} and the related results can be applied to the symmetric case and its special case directly.

\item \textbf{The total forager population $A(t)$:} The population $A$ increases by the maturation rate of brood and decreases with a density-dependent death rate $dA^2$. The density-dependent mortality rate follows the approach of \cite{kang2011mathematical}, thus we have following equation:
\[A' = \underbrace{\beta L}_{\text{maturation from brood}} -\underbrace{dA^2}_{\text{average mortality rate}}\\\]

\item \textbf{The protein forager population $A_p(t)$:} The ratio $\frac{A_p}{A_c}$ measures the nutritional status of the colony. Assume that the brood can survive in a range of nutrient ratio, i.e. $\frac{A_p}{A_c}\in (\theta_0,\theta_c)$, and any ratio greater than $\theta_c$ can be toxic to brood. In addition, there is an optimal nutritional ratio $\frac{A_p}{A_c}$, denoted by $\theta_m$, such that brood could have the optimum survival rate at this ratio. More specifically, the brood survival rate  increases with respect to the value of $\frac{A_p}{A_c}$ when $\frac{A_p}{A_c}\in (\theta_0, \theta_m)$, and passing this optimal ratio $\theta_m$, the brood survival rate decreases with $\frac{A_p}{A_c}$.  The survival rate of brood is zero when $\frac{A_p}{A_c}\leq  \theta_0$ or $\frac{A_p}{A_c}\geq  \theta_c$. Thus, 
\begin{itemize}
\item  \cyan{The portion of the successful brood developed into adults which enter into the protein forager pool can be modeled by the term:}
$\beta L \cdot\max\left\{0,\frac{\partial S_L}{\partial\frac{A_p}{A_c}}\right\},$
\cyan{where $\max\Big\{0,\frac{\partial S_{L}}{\partial \frac{A_p}{A_c}}\Big\}\in [0,1]$ represents the nutritional requirements of the colony measured by the ratio $\frac{A_p}{A_c}$ and a nutrient collection factor.}
\item Based on the nutritional requirements of the colony and other related stimuli, a protein forager can become a carbohydrate forager, and vice versa. This task switching rate depends upon different factors such as the nutritional status of the colony, presence of larvae, individual preference, food type, food concentration and hunger level \cite{cassill1999regulation,portha2002self,pohl2016colony,clark2011behavioral,mailleux2006starvation,dussutour2008carbohydrate}. In this paper, we assume that the task switching rate of protein foragers to carbohydrate foragers depends on the brood population $L$,  the nutritional requirement of the colony $\max\left\{0,\frac{\partial S_L}{\partial\frac{A_c}{A_p}}\right\}$, and the available carbohydrate forager $A_c=A-A_p$, thus its switching rate is $\max\left\{0,\frac{\partial S_L}{\partial{\frac{A_c}{A_p}}}\right\}A_pL$. Similarly, the switching rate of carbohydrate foragers to protein foragers is termed as $\max\left\{0,\frac{\partial S_L}{\partial\frac{A_p}{A_c}}\right\} A_cL$. This gives the net task switching rate of the protein forager: $\left(\max\left\{0,\frac{\partial S_L}{\partial\frac{A_p}{A_c}}\right\}A_c-\max\left\{0,\frac{\partial S_L}{\partial{\frac{A_c}{A_p}}}\right\}A_p\right)L$. {For instance, if the ratio $\frac{A_p}{A_c}$ is less than the optimal nutrient ratio $\theta_m$ (where the maximum brood survival rate occurs), then we expect $\frac{\partial S_L}{\partial\frac{A_p}{A_c}}>0$ and $\frac{\partial S_L}{\partial\frac{A_c}{A_p}}<0$, thus this indicates that carbohydrate foragers will switch tasks to forage for protein, i.e. $\max\left\{0,\frac{\partial S_L}{\partial{\frac{A_c}{A_p}}}\right\}A_pL=0$. In a similar fashion, if the ratio $ \theta_m\leq\frac{A_p}{A_c}\leq \theta_c$, protein foragers will switch to forage for carbohydrates, that is, $\max\left\{0,\frac{\partial S_L}{\partial\frac{A_p}{A_c}}\right\} A_cL=0$.}
\item \cyan{The total forager population $A=A_p+A_c$ decreases with a density-dependent death rate $dA^2=dA(A_p+A_c)$, then the protein forager population decreases with the density-dependent mortality rate $dAA_p$.}\\ 
\end{itemize}
Considering the factors above, we derive the population dynamics of the protein forager as follows:
\small{\[A_p'=\underbrace{\beta L \cdot\max\left\{0,\frac{\partial S_L}{\partial\frac{A_p}{A_c}}\right\}}_{\text{portion of matured adults entering $A_p$}}+\underbrace{\left(\max\left\{0,\frac{\partial S_L}{\partial\frac{A_p}{A_c}}\right\}A_c-\max\left\{0,\frac{\partial S_L}{\partial{\frac{A_c}{A_p}}}\right\}A_p\right)L}_{\text{net task switching}}-\underbrace{{dA A_p}}_{\text{mortality rate}}.\]}
\end{enumerate}

Based on the ecological assumptions above, the population dynamics of a social insect colony with nutrient regulating foraging activities is described as follows:
\begin{equation}\label{M1}
\begin{aligned}
L'&=\gamma S_{L_{\max}}\frac{aA^2}{b+aA^2}-\beta L \\
A' &= \beta L-dA^2 \\
A_p'&=\beta L \cdot\max\left\{0,\frac{\partial S_L}{\partial\frac{A_p}{A_c}}\right\}+
\left(\max\left\{0,\frac{\partial S_L}{\partial\frac{A_p}{A_c}}\right\}A_c-\max\left\{0,\frac{\partial S_L}{\partial{\frac{A_c}{A_p}}}\right\}A_p\right)L-dAA_p  .
\end{aligned}
\end{equation}
The biological meaning of the parameters and the related values are listed in Table \ref{tab1}.

\begin{table}[!ht]
\centering
{\begin{tabular}{lllclcl}
\hline
\textbf{Parameter} &\textbf{Description} & \textbf{Range}  & \textbf{Reference}     \\   \hline
$a$ & Portion of the division of labor invested on larvae & $(0,\>0.25)$ & \cite{kang2011mathematical} \\
$b$ & Half-saturation constant & $(0.001,\>10)$ & \cite{kang2011mathematical} \\
$d$ & Adult worker death rate & $(0.001,\>1)$ &  \cite{kang2011mathematical}\\
$\alpha, \alpha_i, i=1,2$ & Scaling factor(s) of nutrient collection & $(0,1)$ &\\
$\beta$ & Maturation rate from larvae to adult worker & $(0,\>1)$ & \cite{clark2011behavioral} \\
$\gamma$ & Average egg laying rate of queen  & $(0,\>1)$ & \cite{dussutour2008carbohydrate}\\
$\theta _m$ & Optimal nutrient ratio & $(0,\>\theta_c)$ &\\
$\theta_c$ & Maximal nutrient ratio that brood can survive & $(0,\>\frac{1}{\alpha}+\theta_m)$\\
$\theta_0$ & Minimal nutrient ratio for brood survival& $(0, \theta_m)$\\
\hline
\end{tabular}}
\caption{Parameter description and interval values of Model~\eqref{M1}.}
\label{tab1}
\end{table}

\section{Mathematical analysis}\label{mathanalysis}
The state space of the proposed ecological model \eqref{M1} is $\mathbb{R}_+^3$. All parameters  $a,\,b,\,d,\,\alpha,\,\beta,\,\gamma$, $\theta_0, \theta_m,\,\theta_c$ are assumed to be strictly positive based on their biological meaning. We focus on the proposed function $S_L\left(\frac{A_p}{A_c}\right)$ shown in  Eq. \eqref{SLg} and Figure \ref{function-general-1}. The related mathematical results should be easily adopted to the symmetric case \eqref{Sl-der-s}. Under such conditions, we first show that Model \eqref{M1} is biologically well-defined, i.e., it is positively invariant and bounded in $\mathbb{R}_+^3$ in the following lemma:

\begin{lem}\label{invariant}
Model \eqref{M1} is positively invariant and bounded in $\mathbb{R}_+^3=\{(L,A,A_p):L\geq0,\,A\geq0,\,A_p\geq0\}$. In particular, if $L(0)>0,\,A(0)>0$ and $A_p(0)>0$, then $L(t)>0,\,A(t)>0$ and $A_p(t)>0$ for all $t>0$.  
\end{lem}

The extinction equilibrium $E_0=(0,0,0)$ of  Model \eqref{M1} always exists.  The local stability of the trivial equilibrium $E_0$ cannot be analyzed directly for our model \eqref{M1}. However, from the first two equations of Model \eqref{M1}, we have
\begin{equation*} 
\begin{array}{l}
(L+A)'=\gamma S_{L_{\max}}\frac{aA^2}{b+aA^2}-dA^2\leq \big[\frac{\gamma \alpha_1 (\theta_m-\theta_0)a}{b+aA^2}-d\big]A^2\leq[\frac{\gamma \alpha_1 (\theta_m-\theta_0)a}{b}-d]A^2.
\end{array}
\end{equation*} 
Note that Model \eqref{M1} is positively invariant and bounded from Lemma~\ref{invariant}, thus we can conclude that if $\frac{a\alpha_1\gamma  (\theta_m-\theta_0)}{b}<d$, then the inequality above implies that $\limsup_{t\rightarrow\infty} (L+A)$ converges to a nonnegative constant. In addition, we have  $L'|_{A=0,L>0}=-\beta L<0$, $A'|_{L=0}=-d A^2<0$ and ${A_p}'|_{L=0,A>0}=-dA_pA<0$. Therefore, if $\frac{a\alpha_1\gamma (\theta_m-\theta_0)}{b}<d$, then for some initial conditions around $E_0=(0,0,0)$, Model \eqref{M1} converges to the extinction equilibrium $E_0$ in $\mathbb{R}_+^3$. Thus, we have the following proposition:

\begin{prop}\label{Prop-1}
If $\frac{a\alpha_1 \gamma (\theta_m-\theta_0)}{b}<d$, then for some initial condition around the extinction equilibrium point $E_0=(0,0,0)$, taken in $\mathbb{R}_+^3$, the trajectory of Model \eqref{M1} converges to $E_0$.
\end{prop}

\noindent\textbf{Remarks:} Note that the inequality $\frac{a \alpha_1\gamma  (\theta_m-\theta_0)}{b}<d$ implies that  $a<\frac{bd}{\gamma \alpha_1 (\theta_m-\theta_0)}$. Proposition~\ref{Prop-1} implies that if $a$ is not large enough (i.e., the investment to the brood growth is small), or the death rate of adults is too large, then the brood population and the total forager population approaches the extinction equilibrium point $E_0$. In the symmetric case, we have $\theta_0=2\theta_m-\theta_c$, then the inequality becomes  $\frac{a\alpha \gamma (\theta_c-\theta_m)}{b}<d$ with $\alpha_1=\alpha_2=\alpha$.\\

Assume that $E^*=(L^*,A^*,A_p^*)$ is an interior equilibrium of Model \eqref{M1} with the general case of $S_L$. Then based on the equation of $\frac{\mathrm{d}A_p}{\mathrm{d}t}$ shown in \eqref{M1}, we can conclude that $A_p^*$ can exist only if $\frac{\partial S_L}{\partial\frac{A_p}{A_c}}>0$ as it requires $\max\left\{0,\frac{\partial S_L}{\partial\frac{A_p}{A_c}}\right\}>0$. Biologically, this implies that the colony survival requires the nutritional needs of brood being on the positive gradient of the brood survival probability $S_{L_{\max}}$.
Thus, we have $\frac{A_p^*}{A_c^*}=\frac{A_p^*}{A^*-A_p^*}\in (\theta_0,\theta_m)$, and therefore $S_L\left(\frac{A_p}{A_c}\right)=\alpha_1 \left(\frac{A_p}{A_c}-\theta_0\right) $ and 
$$\frac{\partial S_L}{\partial\frac{A_p}{A_c}}\Big|_{A_c=A^*_c,A_p=A_p^*}=\alpha_1;\qquad \frac{\partial S_L}{\partial\frac{A_c}{A_p}}\Big|_{A_c=A^*_c,A_p=A_p^*}=-\alpha_1 \left(\frac{A_p^*}{A_c^*}\right)^2<0.$$

To solve for $(L^*,A^*,A_p^*)$, we set $L'=A'=A'_p=0$, which implies the following equations
\begin{equation*}
\begin{array}{l}
L'=0 \qquad \Longrightarrow  \qquad \alpha_1\gamma\big[\frac{A_p}{A_c}-\theta_0\big]\frac{aA^2}{b+aA^2}-\blue{\beta L}=0,\\[8pt]
A'=0\qquad \Longrightarrow \qquad L=\frac{d}{\beta}A^2,\\[8pt]
A_p'=0\qquad \Longrightarrow \qquad \alpha_1\beta L+\alpha_1 A_cL -dA_p A=0\Longrightarrow L=\frac{dA_pA}{\alpha_1\beta+\alpha_1 A_c} \textcolor{cyan}{,}
\end{array}
\end{equation*}
which gives
\begin{equation*}
 L=\frac{d}{\beta}A^2\qquad \mbox{and} \qquad L=\frac{dA_pA}{\alpha_1\beta+\alpha_1 A_c}.
\end{equation*}
Therefore, $A^*$ of an interior equilibrium $(L^*,A^*,A_p^*)$ satisfies the following equation: 
\begin{equation}\label{root-A}
\begin{array}{l}
ad\beta(1-\alpha_1)A^2-a\alpha_1^2\gamma A+\beta[(1-\alpha_1)(bd+a\alpha_1\gamma\theta_0)-a\alpha_1^2\gamma]=0.\\
\end{array}
\end{equation}
Recall that $\alpha_1\in (0,1)$. Depending on the exact values of $a,\,b,\,d,\,\alpha_1,\,\beta,\,\gamma,\,\theta_0,\,\theta_m,\,\theta_c$, the equation~\eqref{root-A} can have zero, one, or two positive roots. \\
\blue{Let}
\blue{ \[   A_1=\frac{a\alpha_1^2\gamma-\sqrt{{\Delta}}}{2a\beta d(1-\alpha_1)},\quad  
A_2=\frac{a\alpha_1^2\gamma+\sqrt{{\Delta}}}{2a\beta d(1-\alpha_1)},\]
where $${\Delta}=a(a\alpha_1^4\gamma^2-4d\beta^2[(1-\alpha_1)^2(bd+a\alpha_1\gamma\theta_0)-a\alpha_1^2\gamma(1-\alpha_1)])$$ be the possible positive roots of equation \eqref{root-A}.}
Let us denote $\hat{a}^*$ as follows:
\begin{equation}\label{eq-b0-g}
\begin{aligned}
\hat{a}^*&=\frac{4bd^2\beta^2(1-\alpha_1)^2}{\alpha_1^4\gamma^2+4d\beta^2\alpha_1\gamma(1-\alpha_1)[\alpha_1-\theta_0(1-\alpha_1)]}\\\\
&=\frac{bd(1-\alpha_1)}{\alpha_1\gamma(\alpha_1-(1-\alpha_1)\theta_0)}\frac{4d\beta^2(1-\alpha_1)}{\frac{\alpha_1^3\gamma}{\alpha_1-(1-\alpha_1)\theta_0}+4d\beta^2(1-\alpha_1)}<\frac{bd(1-\alpha_1)}{\alpha_1\gamma(\alpha_1-(1-\alpha_1)\theta_0)}\\
\end{aligned}
\end{equation}which is an increasing function of $\theta_0$.\\

In the symmetric case, we have $\theta_0=2\theta_m-\theta_c$, then $\hat{a}^*$ shown in \eqref{eq-b0-g} can be rewritten as
\begin{equation}\label{eq-b0}
\tilde{a}^*=\frac{4bd^2\beta^2(1-\alpha)^2}{\alpha^4\gamma^2+4d\alpha\beta^2\gamma(1-\alpha)[\alpha-(1-\alpha)(2\theta_m-\theta_c)]}<\frac{bd(1-\alpha)}{\alpha\gamma[\alpha-(1-\alpha)(2\theta_m-\theta_c)]}.
\end{equation}
Also note that $\frac{\alpha_1^3\gamma+4d\alpha_1\beta^2(1-\alpha_1)}{4d\beta^2(1-\alpha_1)^2}=\frac{\alpha_1^3\gamma}{4d\beta^2(1-\alpha_1)^2}+\frac{\alpha_1}{1-\alpha_1}>\frac{\alpha_1}{1-\alpha_1}$. Then the following theorem provide conditions for existence of equilibrium solutions of Model \eqref{M1}:

\begin{thm}[Existence of Equilibria]\label{th-existence-general}
For Model \eqref{M1},
\begin{enumerate}
\item If $0<a<\hat{a}^*$ and $\theta_0<\frac{\alpha_1^3\gamma+4d\alpha_1\beta^2(1-\alpha_1)}{4d\beta^2(1-\alpha_1)^2}$, then there is only one trivial equilibrium $E_0=(0,0,0)$ and no other positive equilibrium.  
\item If $a=\hat{a}^*$ and $\theta_0<\frac{\alpha_1^3\gamma+4d\alpha_1\beta^2(1-\alpha_1)}{4d\beta^2(1-\alpha_1)^2}$, then Model \eqref{M1} has two positive equilibria which collapse into one equilibrium $E_*$
  \[E_*=(L_*,A_*,A{_{p_*}})=\left(\frac{d}{\beta}A_*^2,\frac{\alpha_1^2\gamma}{2\beta d(1-\alpha_1)},\frac{\alpha_1\beta A_*+\alpha_1 A_*^2}{\beta+\alpha_1A_*}\right)\] in addition to $E_0=(0,0,0)$.
\item If $a>\frac{bd(1-\alpha_1)}{\alpha_1\gamma(\alpha_1-(1-\alpha_1)\theta_0)}$ and $\theta_0<\frac{\alpha_1}{1-\alpha_1}$, then Model \eqref{M1} has only one positive equilibrium $E_2$
\[ E_2=\left(\frac{d}{\beta}A_2^2, A_2,\frac{\alpha_1\beta A_2+\alpha_1 A_2^2}{\beta+\alpha_1A_2}\right)\] in addition to $E_0$.
\item If $\hat{a}^*<a<\frac{bd(1-\alpha_1)}{\alpha_1\gamma(\alpha_1-(1-\alpha_1)\theta_0)}$ and $\theta_0<\frac{\alpha_1}{1-\alpha_1}$, then Model \eqref{M1} has two positive equilibria in the following form in addition to $E_0$:
\[E_1=\left(\frac{d}{\beta}A_1^2, A_1,\frac{\alpha_1\beta A_1+\alpha_1 A_1^2}{\beta+\alpha_1 A_1}\right) \quad \mbox{and} \quad E_2=\left(\frac{d}{\beta}A_2^2, A_2,\frac{\alpha_1\beta A_2+\alpha_1A_2^2}{\beta+\alpha_1 A_2}\right).\]
\end{enumerate}
\end{thm}

\noindent\textbf{Remarks:} The detailed proof of Theorem \ref{th-existence-general} is shown in the last section. The number of equilibria of Model \eqref{M1} is determined by the positive root(s) of equation \eqref{root-A}. Theorem \ref{th-existence-general} implies that the value of the division of labor invested on larvae $a$ and the minimal protein to carbohydrate ratio $\theta_0$ determine the existence of the interior equilibrium $(L_i,A_i,A_{pi}), i=1, 2$. 

Our simulations (see {Section~\ref{simulation}}) suggest that Model \eqref{M1} has simple dynamics: no limit cycle and only equilibrium dynamics. At the stable equilibrium, the ratio describing the nutritional level of the colony is
\begin{equation}\label{P2Cratio}
\frac{A_p^*}{A_c^*}=\frac{\alpha_1(A^*+\beta)}{\beta(1-\alpha_1)}\in (\theta_0,\theta_c).
\end{equation}
\blue{Equation \eqref{P2Cratio} suggests that the larger the total population of workers investing in nutrient collection is, the higher the ratio of protein to carbohydrates will be, i.e., better nutrient status of the colony}. Notice that $A^*$ depends on $\theta_0$, so $\frac{A_p^*}{A_c^*}$ does as well. \\

Now we discuss stability of the interior equilibrium for Model~\eqref{M1}. {Let $E^*=(L^*,A^*,A_p^*)$ be an arbitrary positive interior equilibrium of Model~\eqref{M1}. The Jacobian matrix associated to Model~\eqref{M1} at equilibrium is:
\begin{equation}\label{Jacobian-0-general}
\begin{array}{l}
J|_{E^*}=\left(
                      \begin{array}{ccc}
                        -\beta & J_{12} & J_{13} \\
                        \beta & -2dA^* & 0 \\
                        J_{31} & J_{32} & J_{33}\\
                      \end{array}
                    \right),
\end{array}
\end{equation}

\begin{equation*}
\begin{array}{l}
J_{12}=\frac{a\alpha_1\gamma A^*[A_p^*(bA^*-aA^{*3}-2bA_p^*)-2b(A^*-A_p^*)^2\theta_0]}{(A^*-A_p^*)^2(b+aA^{*2})^2},\quad\quad  
J_{13}=\frac{a\alpha_1\gamma A^{*3}}{(A^*-A_p^*)^2(b+aA^{*2})}>0,\\[6pt]
J_{31}=\alpha_1(\beta+A^*-A_p^*)>0, \quad\quad 
J_{32}=\alpha_1 L^*-dA_p^*,\quad\quad
J_{33}= -(\alpha_1L^*+d A^*)<0.
\end{array}
\end{equation*}
Then the characteristic equation of $J|_{E^*}$ is
\begin{equation}\label{characteristic-0-general}
f(\lambda)=\lambda^3+C_1\lambda^2+C_2\lambda+C_3=0,
\end{equation}
where
\begin{equation}\label{J-lambda-0-general}
\begin{array}{l}
C_1=\beta+\alpha L^*+3d A^*>0,\\[8pt]
C_2=J_{11}J_{33}+J_{11}J_{22}+J_{22}J_{33}-J_{21}J_{12}-J_{31}J_{13},\\[8pt]
C_3=-\textrm{det}(J|_{E_i^*})=J_{11}J_{22}J_{33}+J_{21}J_{32}J_{13}-J_{21}J_{12}J_{33}-J_{31}J_{22}J_{13}.
\end{array}
\end{equation}}

The stability of the steady state $E^*=(L^*,A^*,A_p^*)$ can be determined by the distribution of the roots of Eq.~\eqref{characteristic-0-general}. That is, if all the roots of Eq.~\eqref{characteristic-0-general} have negative real parts, then $E^*$ is locally asymptotically stable; if at least one root of Eq.~\eqref{characteristic-0-general} has positive real parts, then $E^*$ is unstable; if any root has zero real part and other roots all have negative real parts, then the stability of $E^*$ cannot be determined by the linearized system directly.\\

The following theorem provides a global result on dynamics of the proposed model~\eqref{M1} regarding when a colony will collapse.

\begin{thm}[Extinction of species]\label{extinction-E0-general}
If $0<a<\hat{a}^*$ and $\theta_0<\frac{\alpha_1^3\gamma+4d\alpha_1\beta^2(1-\alpha_1)}{4d\beta^2(1-\alpha_1)^2}$, then Model \eqref{M1} has global stability at $E_0=(0,0,0)$.
\end{thm}
\noindent\textbf{Biological implications:} Theorem~\ref{extinction-E0-general} has stronger result than results stated in Proposition \ref{Prop-1}, and indicates that the portion of the division of labor invested on larvae $a$ and the nutrient $\theta_0$ are important factors determining whether larvae and adult worker ants can survive. This theorem provides a sufficient condition leading to the collapse of the colony.

\begin{thm}[Stability Conditions]\label{th-Survival-general} For Model~\eqref{M1},
\begin{enumerate}
\item  Assume that $a>\frac{bd(1-\alpha_1)}{\alpha_1\gamma(\alpha_1-(1-\alpha_1)\theta_0)}$ and $\theta_0<\frac{\alpha_1}{1-\alpha_1}$, then Model~\eqref{M1} has a unique interior equilibrium 
$E_2=(L_2,A_2,A_{p2})=\left(\frac{d}{\beta}A_2^2, A_2,\frac{\alpha_1\beta A_2+\alpha_1 A_2^2}{\beta+\alpha_1A_2}\right)$. If it satisfies $C_1(E_2)C_2(E_2)>C_3(E_2)>0$, then $E_2$ is locally asymptotically stable.
\item Assume that $\hat{a}^*<a<\frac{bd(1-\alpha_1)}{\alpha_1\gamma(\alpha_1-(1-\alpha_1)\theta_0)}$ and $\theta_0<\frac{\alpha_1}{1-\alpha_1}$, Model~\eqref{M1} has two interior equilibria $E_i= (L_i,A_i,A_{pi})=\left(\frac{d}{\beta}A_i^2, A_i,\frac{\alpha_1\beta A_i+\alpha_1A_i^2}{\beta+\alpha_1A_i}\right),\,i=1,2$, where $E_1<E_2$, if $C_1(E_1)C_2(E_1)-C_3(E_1)<0$ but $C_1(E_2)C_2(E_2)>C_3(E_2)>0$, then the interior equilibrium $E_2$ is locally asymptotically stable while $E_1$ is unstable.
\end{enumerate}
\end{thm}

\begin{figure}[!ht]
\centering
\subfigure[$\theta_0=0.1$]
{\includegraphics[scale=0.38]{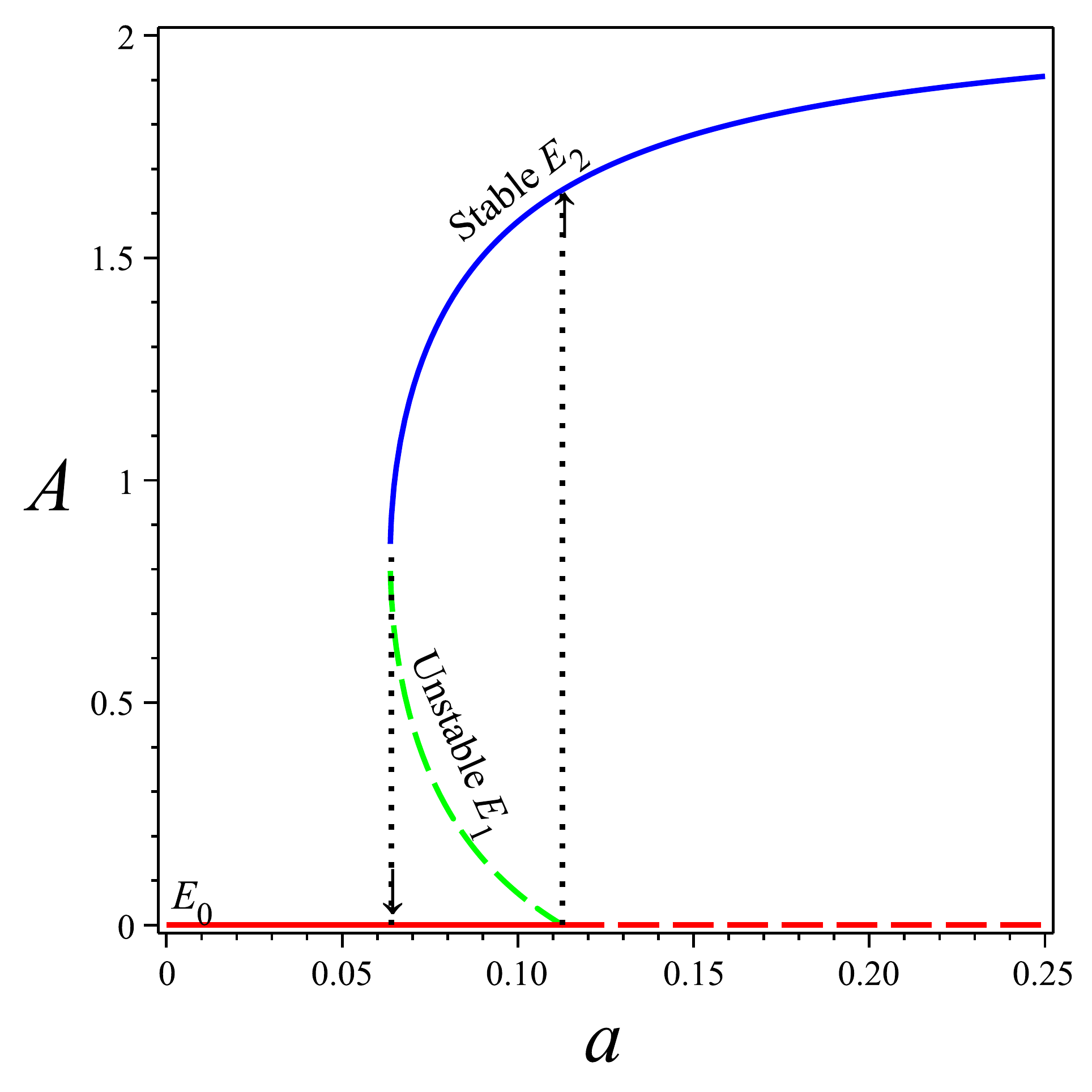}\label{figgeneral-Aa-1} }
\subfigure[$\theta_0=0.2$]
{\includegraphics[scale=0.38]{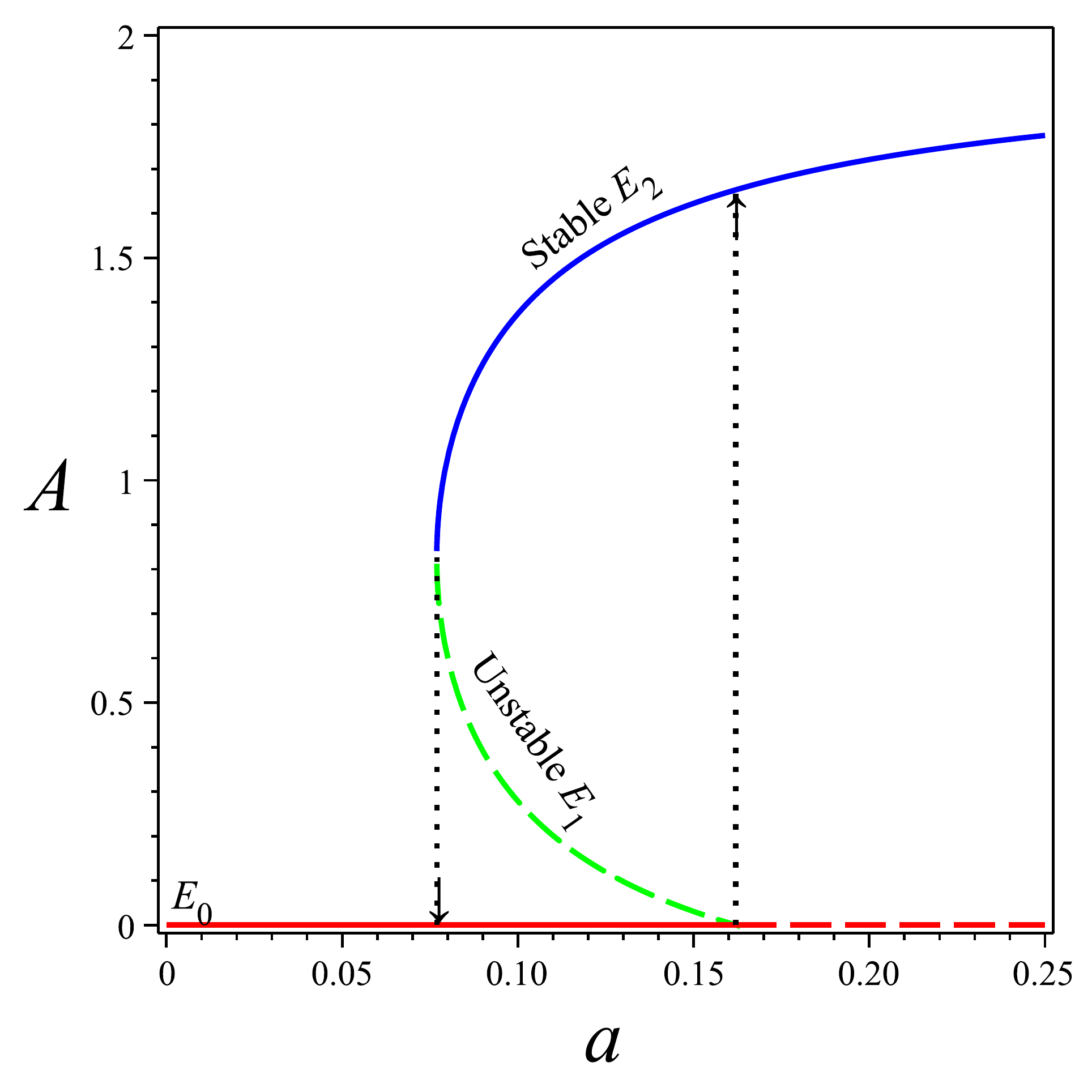}\label{figgeneral-Aa-2} }
\caption{Backward bifurcation diagram of the division of labor invested on brood $a$ v.s. the total forager population $A$. Other parameters  values are $b=0.1,\,d=0.1,\,{\alpha_1=0.3},\,\beta=0.7,\,\gamma=0.9$. The solid line indicates that the equilibrium is locally asymptotically stable while the dashed line indicates that the equilibrium is unstable. The first vertical dotted line is the critical point $\hat{a}^*$ for saddle node bifurcation and the second dotted line is the transition point when the system has two interior equilibrium to one interior equilibrium. The blue color indicates $E_2$ which is always stable; the green color indicates $E_1$ which is always unstable; and the red color is the extinction equilibrium $E_0$. }\label{figgeneral} 
\end{figure}

\begin{figure}[!ht]
\centering
\subfigure[Effects on $\theta_0$]
{\includegraphics[scale=0.38]{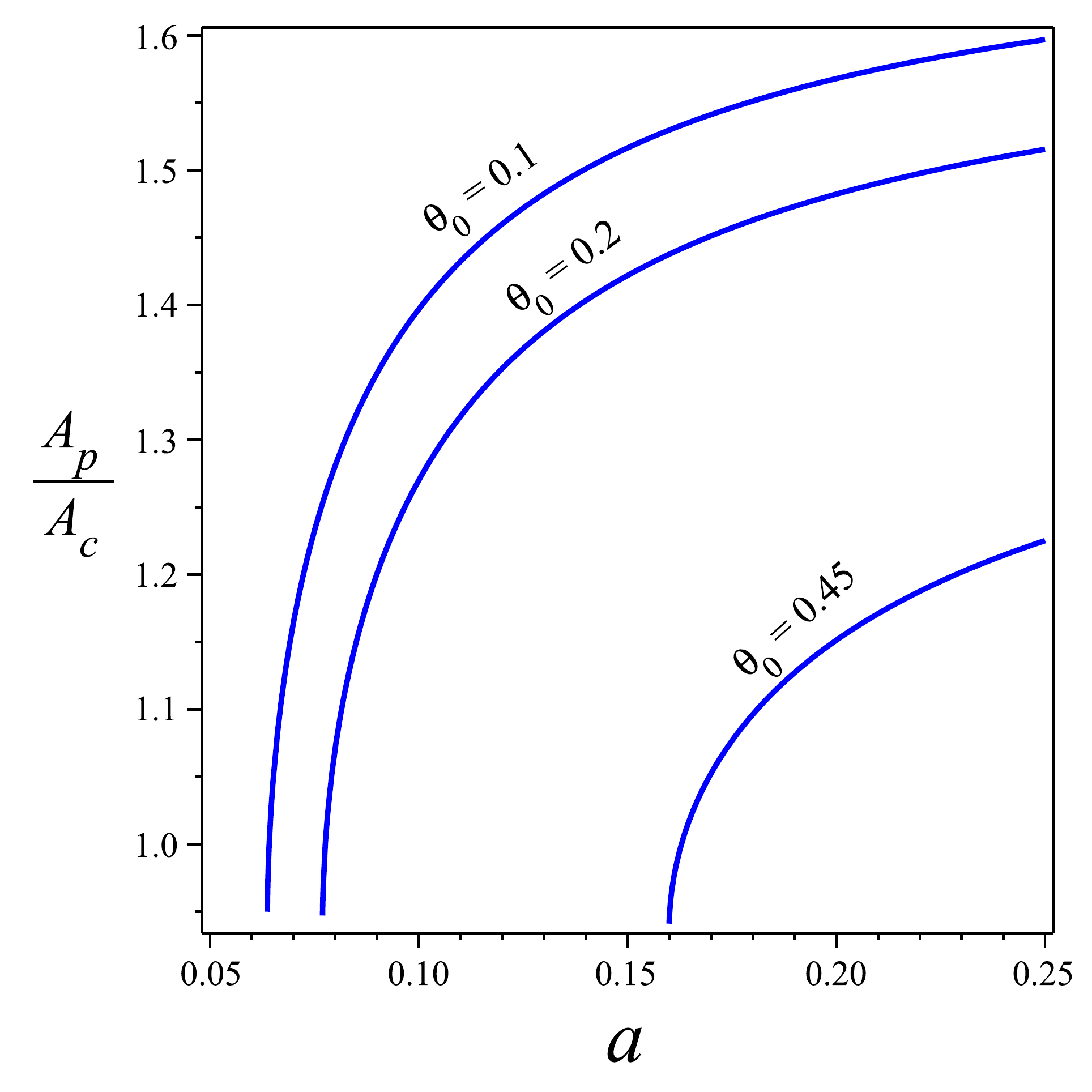}\label{fig-ApAc-a-1} }
\subfigure[Effects on $\theta_m$]
{\includegraphics[scale=0.38]{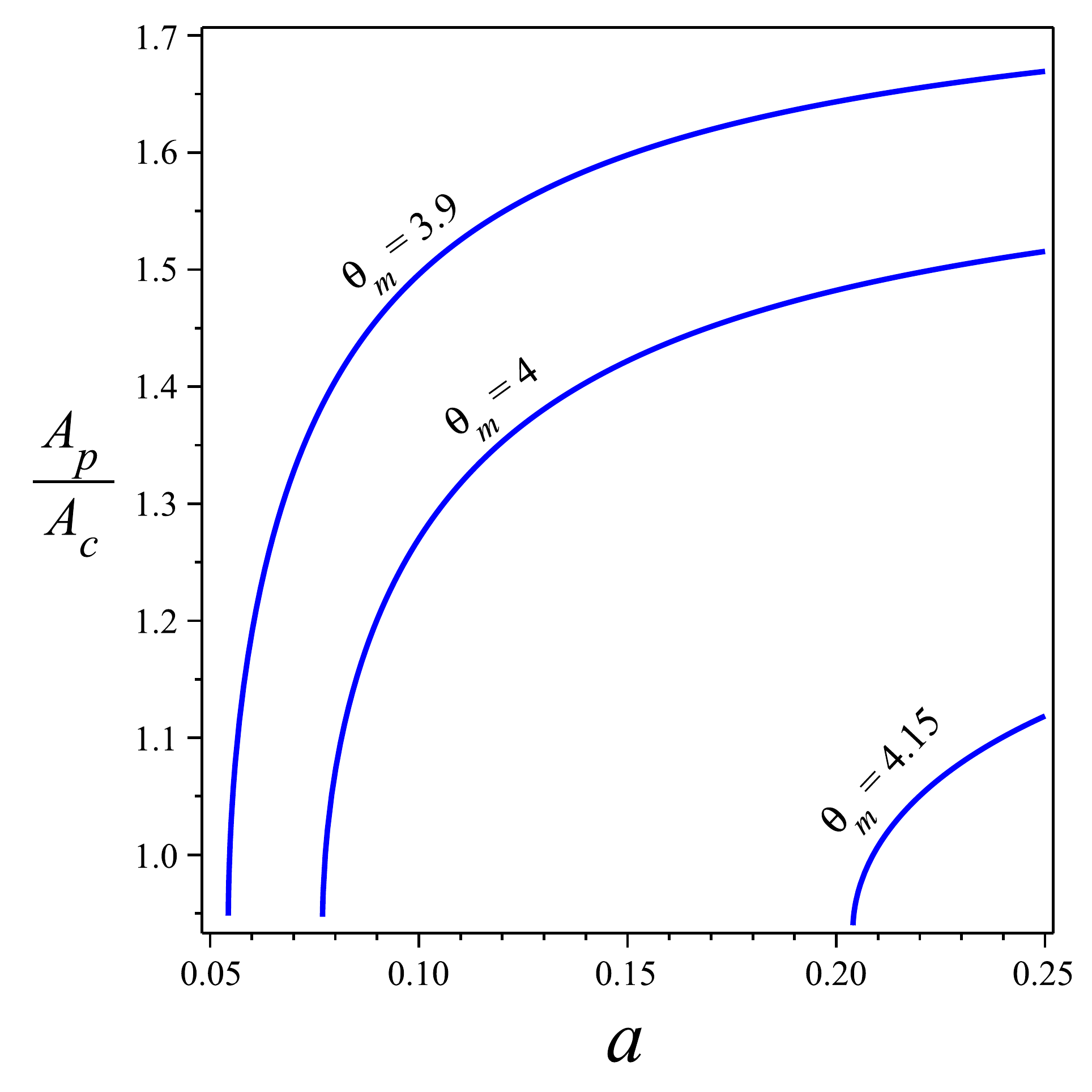}\label{fig-ApAc-a-2} }
\vspace{-10pt}
\caption{Bifurcation diagrams of the ratio of $\frac{A_p}{A_c}$ v.s. the division of labor invested on larvae $a$ with: (a) different values of nutrient threshold $\theta_0$ and (b) different values of optimal nutrient ratio $\theta_m$ when $\theta_c=7.8$. Other parameters values are $b=0.1,\,d=0.1,\,{\alpha_1=0.3},\,\beta=0.7,\,\gamma=0.9$. }\label{fig-ApAc-a}
\end{figure}

\noindent\textbf{Biological Implications:} The results in Lemma \ref{invariant}, Theorems~\ref{th-existence-general} and~\ref{th-Survival-general}, imply that the division of labor invested on larvae $a$ decreases past the critical point $\hat{a}^*=\frac{4bd^2\beta^2(1-\alpha_1)^2}{\alpha_1^4\gamma^2+4d\alpha_1\beta^2\gamma(1-\alpha_1)(\alpha_1-(1-\alpha_1)\theta_0)}$ shown in \eqref{eq-b0-g} and the first dotted line in Figure~\ref{figgeneral}. Model \eqref{M1} exhibits a backward bifurcation shown in Figure~\ref{figgeneral} where $b=0.1,\,d=0.1,\,{\alpha_1=0.3},\,\beta=0.7,\,\gamma=0.9$. In  Figure~\ref{figgeneral-Aa-1}, we set $\theta_0=0.1$ and in Figure~\ref{figgeneral-Aa-2}, we set $\theta_0=0.2$. Based on the expression of the critical value $\hat{a}^*$ shown in \eqref{eq-b0-g}, $\hat{a}^*$ is an increasing function of $\theta_0$, which is reflected in the difference between Figure~\ref{figgeneral-Aa-1} and Figure~\ref{figgeneral-Aa-2}. The value of $\theta_0$ measures the minimum ratio of protein to carbohydrates that can allow the survival of larvae. Simulations shown in Figure~\ref{figgeneral-Aa-1},~\ref{figgeneral-Aa-2} and~\ref{fig-ApAc-a} suggest that the larger value of $\theta_0$, the more likely the colony can survive with a larger population of workers $A$ and thus the higher nutrient ratio $\frac{A_p}{A_c}$. In summary, our theoretical work combined with the related simulations suggest that the division of labor invested on larvae $a$ and the minimal nutrient ratio $\theta_0$ can affect colony survival, the distribution of the brood and the total forager population affect the protein forager population. \blue{For instance, the larger $\theta_0$, the more division of labor invested on larvae is required to ensure survival of the colony. Also, under this scenario, the population distribution of brood and workers is smaller.} {Moreover, Figure~\ref{fig-ApAc-a} shows the bifurcation diagrams of the ratio of $\frac{A_p}{A_c}$ versus the division of labor invested on larvae $a$ with different values of $\theta_0$, other parameters values are taken as those in Figure~\ref{figgeneral}}.  

All our theoretical results can apply to the symmetric case when $\alpha_1=\alpha_2=\alpha$ and $\theta_0=2\theta_m-\theta_c$.  Now we focus on the special case of the symmetric case $\theta_0=0$. For convenience, let
\begin{equation}\label{Eq-a-0}
a^*\triangleq\frac{4bd^2\beta^2(1-\alpha)^2}{\alpha^2\gamma[\alpha^2\gamma+4d\beta^2(1-\alpha)]}=\frac{bd(1-\alpha)}{\alpha^2\gamma}\frac{4d\beta^2(1-\alpha)}{[\alpha^2\gamma+4d\beta^2(1-\alpha)]}<\frac{bd(1-\alpha)}{\alpha^2\gamma}.\\
\end{equation}

Let $\Delta=a^2\alpha^2\gamma[\alpha^2\gamma+4d\beta^2(1-\alpha)]-4abd^2\beta^2(1-\alpha)^2$ and \[   A_1=\frac{a\alpha^2\gamma-\sqrt{\Delta}}{2a\beta d(1-\alpha)},\quad  A_2=\frac{a\alpha^2\gamma+\sqrt{\Delta}}{2a\beta d(1-\alpha)}.\]
  \[(L_*,A_*,{A_p}_*)=\left(\frac{d}{\beta}A_*^2,\frac{\alpha^2\gamma}{2\beta d(1-\alpha)},\frac{\alpha\beta A_*+\alpha A_*^2}{\beta+\alpha A_*}\right).\] 
Define $a_1, a_2$ and $M$ as follows:
\begin{equation}\label{a12-M}
\begin{array}{l} 
a_1=\frac{b\big(\alpha^2\gamma+4d\beta^2(1-\alpha)-\alpha\sqrt{\gamma(\alpha^2\gamma+8d\beta^2(1-\alpha))}\big)}{2\alpha^2\beta^2\gamma},\quad a_2=\frac{b\big(\alpha^2\gamma+4d\beta^2(1-\alpha)+\alpha\sqrt{\gamma(\alpha^2\gamma+8d\beta^2(1-\alpha))}\big)}{2\alpha^2\beta^2\gamma},\\[4pt]
M=\frac{1}{\beta^2}(2\alpha d{A_2}^2+(\alpha+2d)\beta A_2+3\beta^2)+\frac{d}{a\alpha\gamma}\big(\frac{A_2}{{A_p}_2}(a{A_2}^2-b)+2b\big).
\end{array}
\end{equation}
 
\begin{thm}[Dynamics of the Special Symmetric Case]\label{th-Survival-new}
Model \eqref{M1} is positive invariant in $\mathbb{R}_+^3$ and every trajectory attracts to a compact set $\mathbb{C}=\Big[0,\frac{\alpha\gamma\theta_m}{\beta}\Big]\times\Big[0,\frac{\sqrt{d\alpha\gamma\theta_m}}{d}\Big]\times\Big[0,\frac{\alpha\beta\sqrt{d\alpha\gamma\theta_m}+\alpha^2\gamma\theta_m}{d\beta}\Big]$. In addition,
\begin{enumerate}
\item If $a<\min\big\{a^*,\frac{bd}{\alpha\gamma\theta_m}\big\}$, then Model \eqref{M1} has global stability at $E_0=(0,0,0)$.

\item If $a>\frac{bd(1-\alpha)}{\alpha^2\gamma}$, $a_1<a<a_2$ and 
\begin{equation*} 
\begin{array}{l} 
\max\big\{0,\frac{\beta}{\beta-\alpha A_2}\big\}<\frac{A_2}{{A_c}_2}<\min\left\{\frac{\alpha L_2+dA_2}{\alpha L_2+d{A_p}_2}\big[\frac{d(a{A_2}^2-b)}{a\alpha\gamma}+\frac{2{A_p}_2}{A_2}(\frac{bd}{a\alpha\gamma}+1)\big],M\right\},
\end{array}
\end{equation*}
then Model~\eqref{M1} has a unique interior equilibrium  $E_2=(L_2,A_2,{A_p}_2)$ that is locally asymptotically stable.
\item If $a^*<a<\frac{bd(1-\alpha)}{\alpha^2\gamma}$, $a_1<a<a_2$, and 
\begin{equation*} 
\begin{array}{l} 
\frac{A_1}{{A_c}_1}>\frac{1}{\beta^2}(2\alpha d{A_1}^2+(\alpha+2d)\beta A_1+3\beta^2)-\frac{d}{a\alpha\gamma}\big(\frac{A_1}{{A_p}_1}(b-a{A_1}^2)-2b\big), \\[4pt]
\frac{A_2}{{A_c}_2}<\frac{1}{\beta^2}(2\alpha d{A_2}^2+(\alpha+2d)\beta A_2+3\beta^2)+\frac{d}{a\alpha\gamma}\big(\frac{A_2}{{A_p}_2}(a{A_2}^2-b)+2b\big),
\end{array}
\end{equation*}
then Model~\eqref{M1} has two interior equilibria $E_i= (L_i,A_i,{A_p}_i)=\left(\frac{d}{\beta}A_i^2, A_i,\frac{\alpha\beta A_i+\alpha A_i^2}{\beta+\alpha A_i}\right),\,i=1,2$  where the interior equilibrium $E_2$ is locally asymptotically stable while $E_1$ is unstable.
\item If $a>\frac{bd(1-\alpha)}{\alpha^2\gamma}$, $a_1<a<a_2$ and 
\begin{equation*} 
\begin{array}{l} 
\max\left\{0,\frac{\beta}{\beta-\alpha A},\frac{\beta+\alpha A}{\beta(1-\alpha)}\right\}<\frac{A}{A_c}<\min\left\{\frac{\alpha L+dA}{\alpha L+dA_p}\big[\frac{d(aA^2-b)}{a\alpha\gamma}+\frac{2A_p}{A}(\frac{bd}{a\alpha\gamma}+1)\big],M,\frac{adA^2+bd+a\alpha\gamma}{a\alpha\gamma}\right\},
\end{array}
\end{equation*} 
 then Model~\eqref{M1} has a unique interior equilibrium $E_2$ which is globally stable. 
\end{enumerate}
\end{thm}

\noindent\textbf{Remarks:}  Theoretical results and numerical simulations (see {Section~\ref{simulation}}) confirm that the special symmetric case of Model~\eqref{M1}, i.e., $\theta_0=0$ and $\theta_c=2\theta_m$, undergoes a backward bifurcation as $a$ decreases past the critical value $a^*$ defined in \eqref{Eq-a-0}. Conditions shown in Theorem \ref{th-Survival-new} suggest the importance of the protein to carbohydrate ratio, i.e., $\frac{A_2}{{A_c}_2}=1+\frac{{A_p}_2}{{A_c}_2}$, in determining the colony population dynamics. \\ 

\noindent\textbf{Summary of Dynamics:} According to our analytical results shown in this section, we can conclude that Model~\eqref{M1} undergoes a backward bifurcation as $a$ decreases past $\hat{a}^*$ (or ${a}^*$ in the case of $\theta_0=0$ and $\theta_c=2\theta_m$). More specifically, it exhibits the following global dynamics:  
\begin{enumerate}
\item If $0<a<\hat{a}^*$ and $\theta_0<\frac{\alpha_1^3\gamma+4d\alpha_1\beta^2(1-\alpha_1)}{4d\beta^2(1-\alpha_1)^2}$, then the colony collapses due to the lack of efforts of division of labor invested on larvae and the minimum nutrient requirement $\theta_0$ being too low.
\item If $0<\hat{a}^*<a<\frac{bd(1-\alpha_1)}{\alpha_1\gamma(\alpha_1-(1-\alpha_1)\theta_0)}$, the survival of the colony depends on its initial population size.
\item If $a>\frac{bd(1-\alpha_1)}{\alpha_1\gamma(\alpha_1-(1-\alpha_1)\theta_0)}>0$, then the colony persists.
\end{enumerate}
Our theoretical results suggest that the survival rate of larva to worker ${S_{L_{\max}}}\left(\frac{A_p}{A_c}\right)$  plays critical roles in determining colony population dynamics. We assume that ${S_{L_{\max}}}\left(\frac{A_p}{A_c}\right)$ takes the form of \eqref{SLg} based on relevant biological studies. Our analysis implies that the values of parameters $\alpha_1$ and $\theta_0$ in ${S_{L_{\max}}}\left(\frac{A_p}{A_c}\right)$ have pronounced impacts on dynamical outcomes. In the next section, we use bifurcation diagrams to explore detailed impacts. 

\section{Numerical simulations}\label{simulation}

In this section, we use numerical simulations to illustrate equilibrium dynamics of the proposed  model and obtain further biological insights on the dynamical outcomes of certain life history parameters of the colony. 

For the general case of ${S_{L_{\max}}}\left(\frac{A_p}{A_c}\right)$, the dynamics of Model~\eqref{M1} depends on the division of labor $a$, egg laying rate $\gamma$, the scaling factor on the brood survival rate due to the nutritional status $\alpha_1$, the minimal nutrition ratio $\theta_0$, the maturation rate $\beta$, and the natural mortality $d$. To explore the effects of $a$ and $\theta_0$, we perform bifurcation diagrams in Figure~\ref{figgeneral} and Figure~\ref{fig-ApAc-a} by setting 
$$b=0.1,\,d=0.1,\,{\alpha_1=0.3},\,\beta=0.7,\,\gamma=0.9.$$

Figure~\ref{figgeneral} and Figure~\ref{fig-ApAc-a} suggest that (1) small values of division of labor $a$ can lead to colony collapse; (2) intermediate values of $a$ can make the system go through saddle node bifurcation; and (3) large values of $a$ can insure colony survival. This implies that Model~\eqref{M1} goes through backward bifurcation on $a$.  We can see that the larger value of $a$ can lead to the larger population $A$ (see Figure~\ref{figgeneral}) and the larger nutrient ratio $\frac{A_p}{A_c}$ (see Figure~\ref{fig-ApAc-a}). Figure~\ref{figgeneral} and~\ref{fig-ApAc-a} also show the effects of the minimal nutrient requirement for brood survival $\theta_0$: The larger value of $\theta_0$, (1) the larger critical threshold  $\hat{a}^*$; (2) the smaller population $A$; and (3) the smaller nutrient status, i.e., the smaller value of  $\frac{A_p}{A_c}$.

\begin{figure}[!ht]
\centering
\subfigure[Effects of $\theta_0$ on population dynamics]
{\includegraphics[scale=0.35]{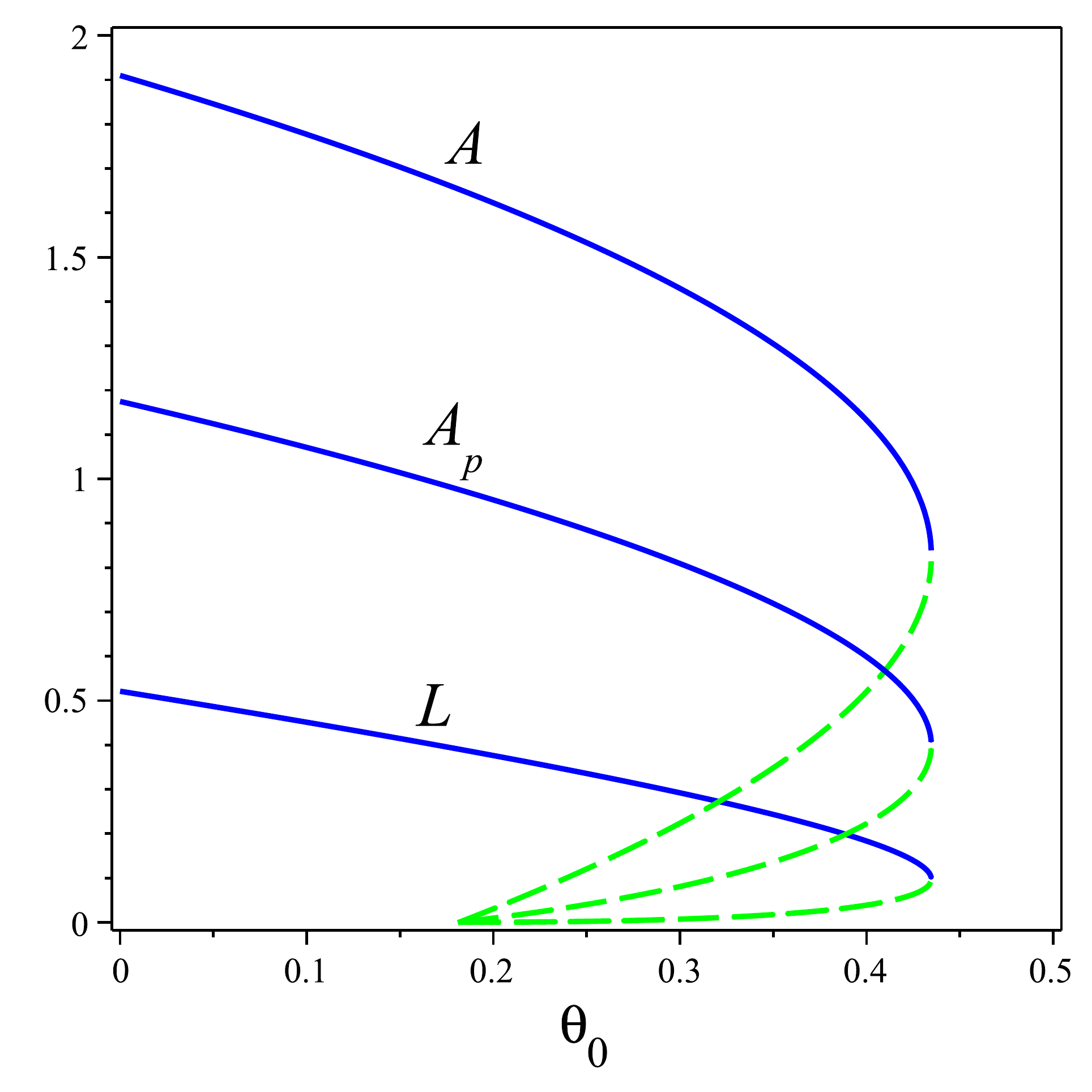}\label{fig-LAAp-theta0}}
\subfigure[Effects of $\alpha_1$ on $A$]
{\includegraphics[scale=0.35]{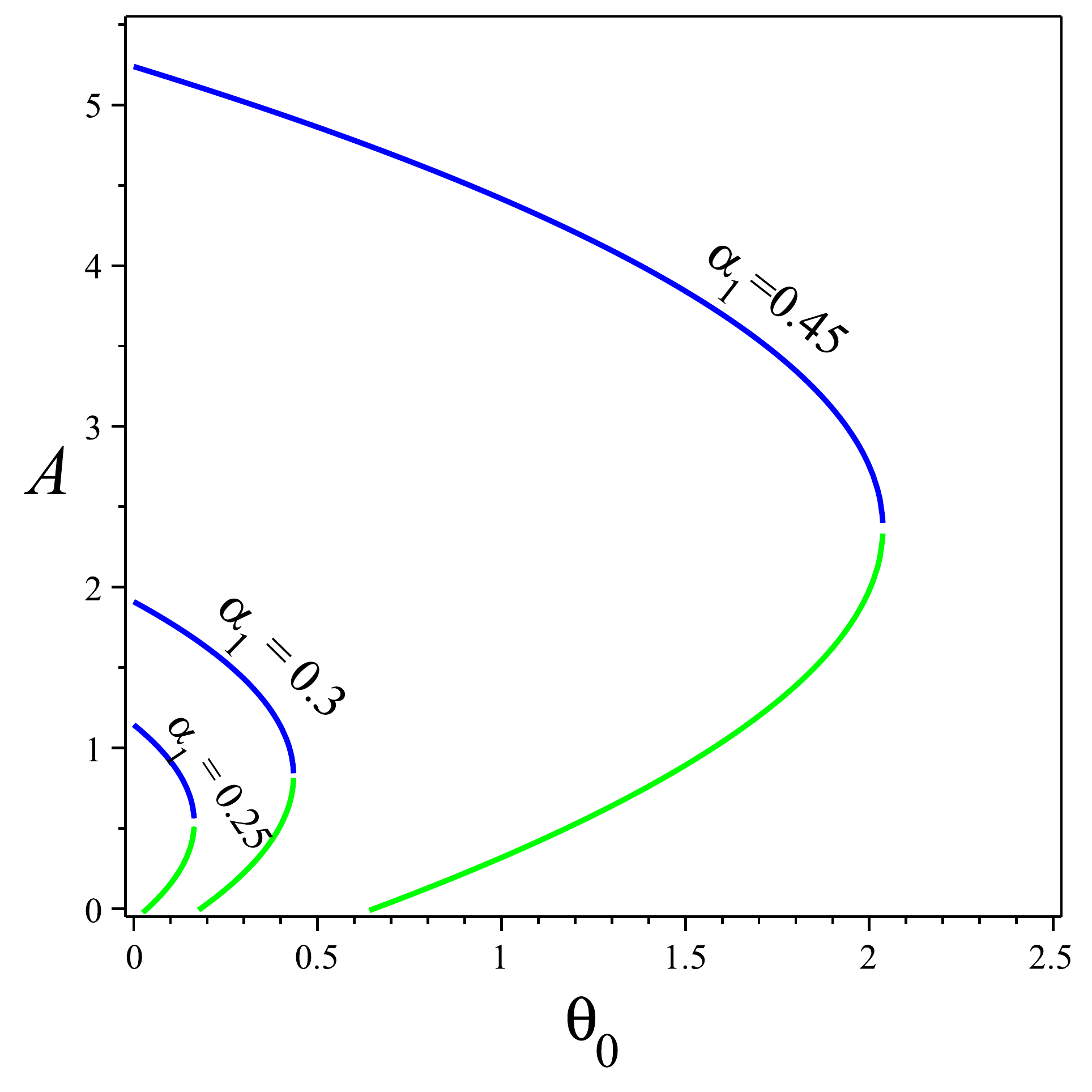}\label{fig-A-a-alpha}}
\caption{{(a) The bifurcation diagrams of $L,\,A$ and $A_p$ v.s. the nutrient threshold $\theta_0$ with $\alpha_1=0.3$; and (b) the bifurcation diagram of $A$  v.s. the nutrient threshold $\theta_0$ with different values of $\alpha_1$. Other parameters values are taken as $a=0.15,\,b=0.1,\,d=0.1,\,\beta=0.7,\,\gamma=0.9$. }}\label{fig-LAAp-theta0-alpha}
\end{figure}

Next, we perform bifurcation diagrams of Model \eqref{M1} regarding how the minimum nutritional requirement $\theta_0$ and the scaling factor of survival probability of brood  $\alpha_1$ affect population dynamics of the colony in Figure \ref{fig-LAAp-theta0-alpha}. Figure \ref{fig-LAAp-theta0} shows that Model \eqref{M1} exhibits reversed backward bifurcation on $\theta_0$. Figure \ref{fig-A-a-alpha} suggests that the larger value of $\alpha_1$, the larger population of worker $A$ and the better probability of colony survival.

In the remaining of this section, we focus on the symmetric case of ${S_{L_{\max}}}\left(\frac{A_p}{A_c}\right)$ shown in \eqref{Sl-der-s} where $\alpha_1=\alpha_2=\alpha$ and $\theta_0=2\theta_m-\theta_c$.

\begin{figure}[!ht]
\centering
\subfigure[$2\theta_m=\theta_c=8$]
{\includegraphics[width=70mm]{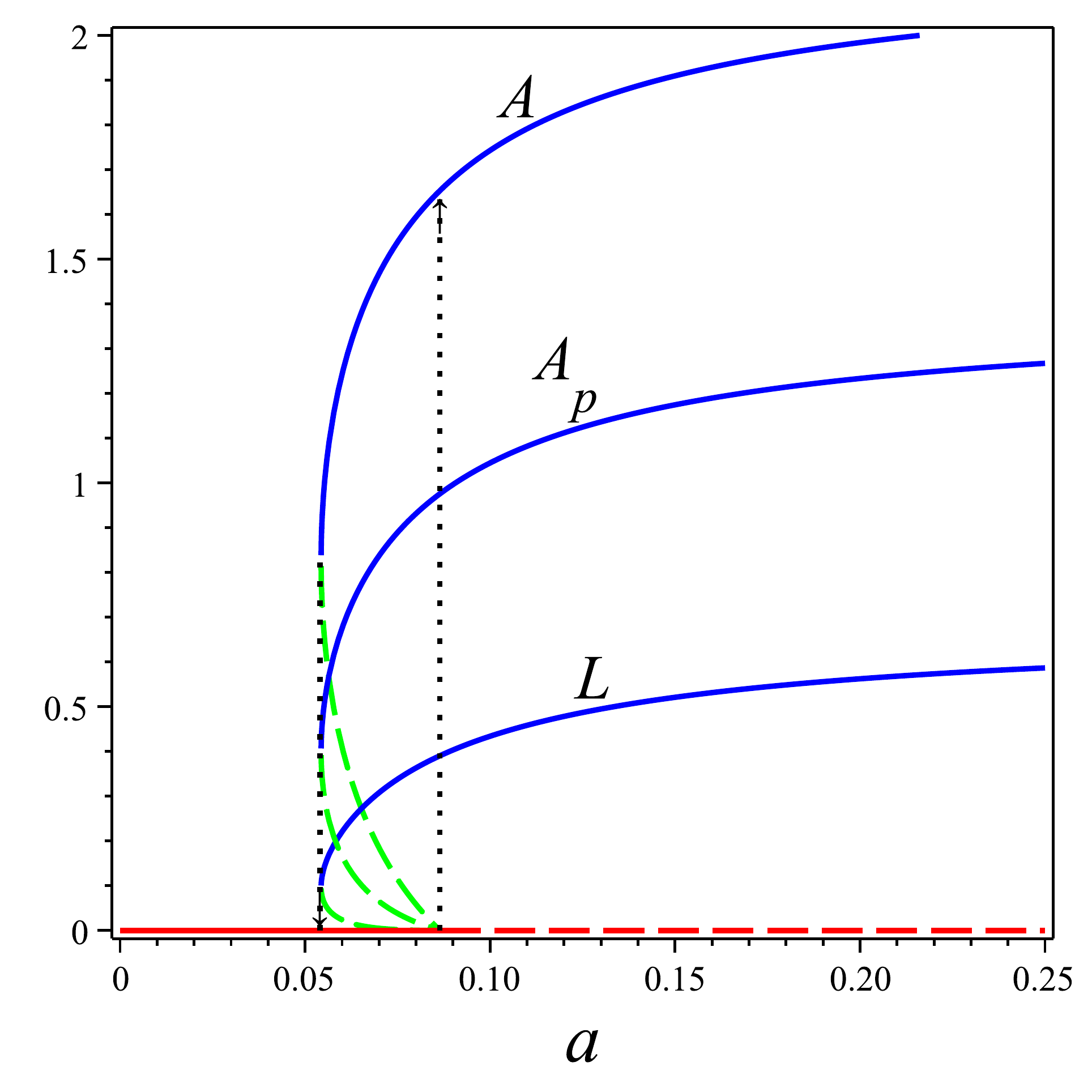}\label{Figbif-a}}  
\subfigure[$2\theta_m>\theta_c=\textcolor{black}{7.8}$]
{\includegraphics[width=70mm]{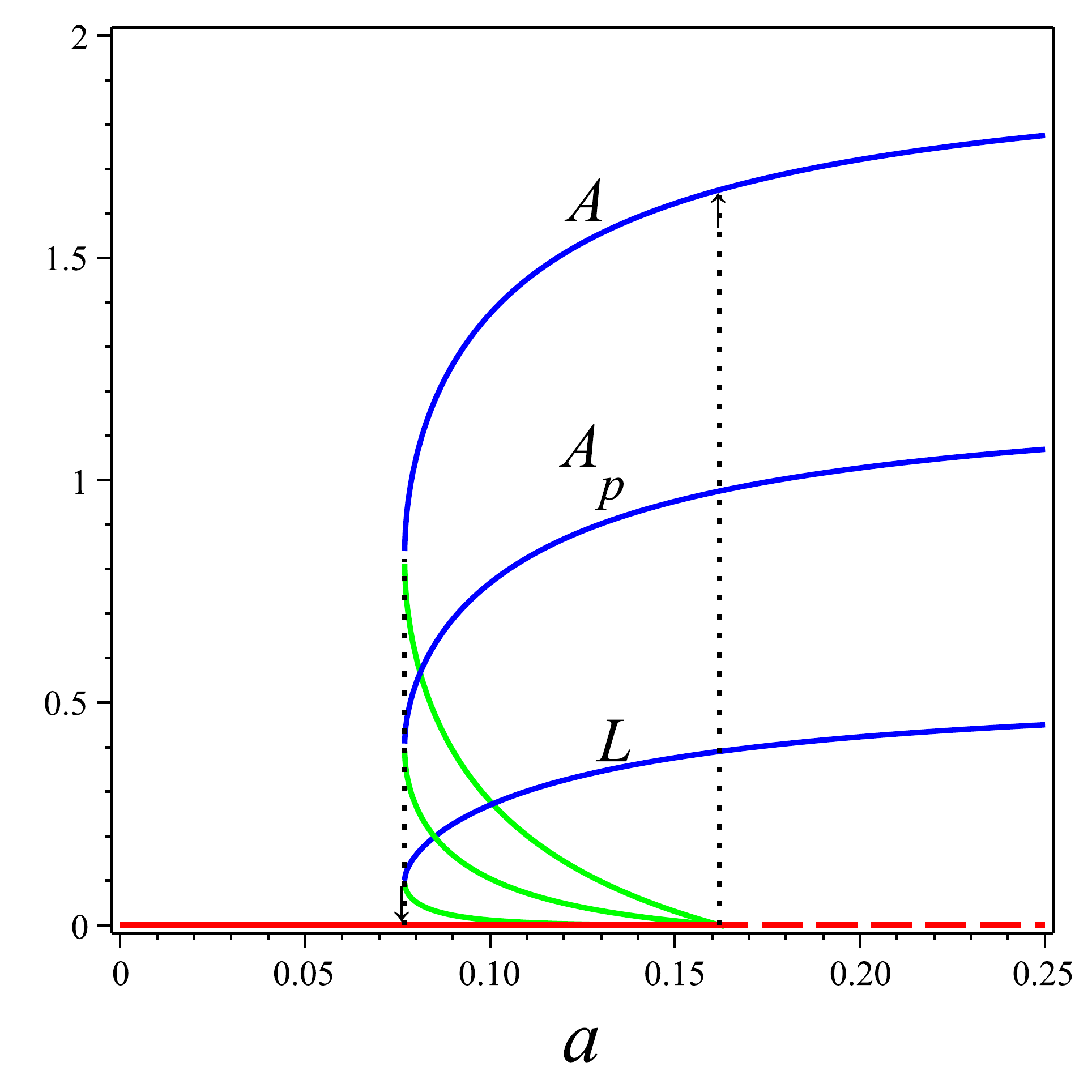}\label{Figbif-Aa}} 
\caption{Bifurcation diagrams of the division of labor invested on larvae $a$ for Model~\eqref{M1}. Backward bifurcation occurs at $a^*\triangleq\frac{4bd^2\beta^2(1-\alpha)^2}{\alpha^2\gamma[\alpha^2\gamma+4d\beta^2(1-\alpha)]}$ for the case of $2\theta_m=\theta_c=8$ (Figure \ref{Figbif-a}); and at ${\tilde{a}^*=\frac{4bd^2\beta^2(1-\alpha)^2}{\alpha^4\gamma^2+4d\alpha\beta^2\gamma(1-\alpha)(\alpha-(1-\alpha)(2\theta_m-\theta_c))}}$ for the case of $2\theta_m>\theta_c=7.8$ (Figure \ref{Figbif-Aa}). Other parameters values are taken as $b=0.1,\,d= 0.1,\,\alpha=0.3,\,\beta = 0.7,\,\gamma = 0.9,\,\theta_m=4$. The solid line indicates that the equilibrium is locally asymptotically stable while the dotted line indicates that the equilibrium is unstable. The blue color indicates $E_2$ which is always stable; the green color indicates $E_1$ which is always unstable; and the red color is the extinction equilibrium $E_0$. }
\end{figure}

 \textbf{Special symmetric case $\theta_c=2\theta_m$ (i.e., $\theta_0=0$)}: Figure~\ref{Figbif-a} provides an example of bifurcation diagram on division of labor invested on larvae $a$ of Model~\eqref{M1} by choosing the following parameters values:  
\[b=0.1,\,d= 0.1,\,\alpha=0.3,\,\beta = 0.7,\,\gamma = 0.9,\,\theta_c = 8,\,\theta_m=4.\]

\noindent Figure~\ref{Figbif-a} shows that Model~\eqref{M1} goes through backward bifurcation at $a^*=\frac{4bd^2\beta^2(1-\alpha)^2}{\alpha^2\gamma[\alpha^2\gamma+4d\beta^2(1-\alpha)]}=0.054$. If $a<a^*$, then colony collapses; if $0.054=a^*<a<\frac{bd(1-\alpha)}{\alpha^2\gamma}$, Model~\eqref{M1} has two positive interior equilibria $E_1=(L_1,A_1,{A_p}_1)$ and $E_2=(L_2,A_2,{A_p}_2)$ with $E_2$ being locally stable; and if $a>\frac{bd(1-\alpha)}{\alpha^2\gamma}$, then the colony survives.

\begin{figure}[!ht]
\centering
\subfigure[{Time series of Model~\eqref{M1} when $\theta_c=2\theta_m$ when the portion of the division of labor invested on larvae is $a=0.07$.}]
{\includegraphics[width=80mm]{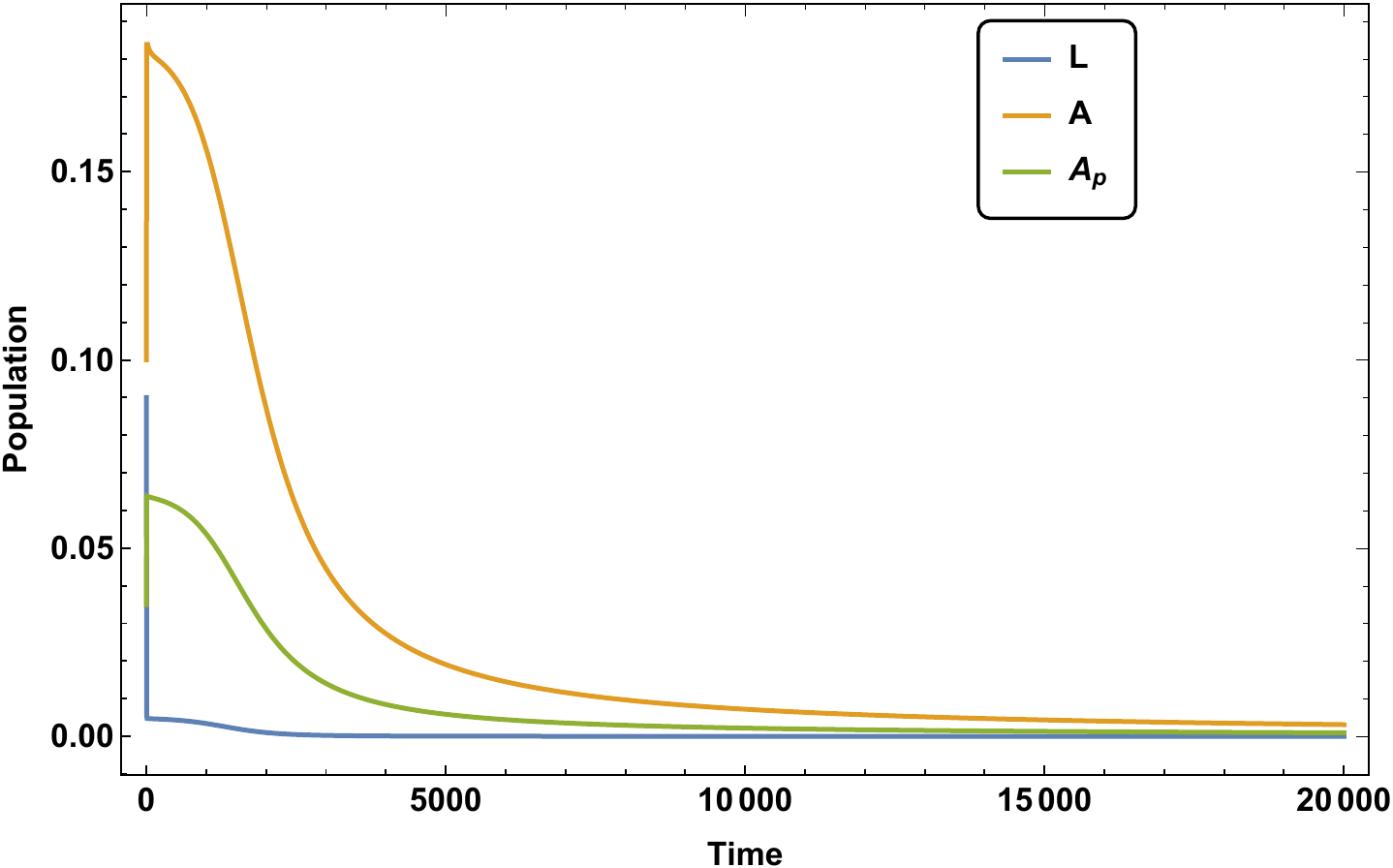}\label{time-1-1}} \quad
\subfigure[{The time series of Model~\eqref{M1} when $\theta_c=2\theta_m$ when the portion of the division of labor invested on larvae is $a=0.1$.}  ]
{\includegraphics[width=80mm]{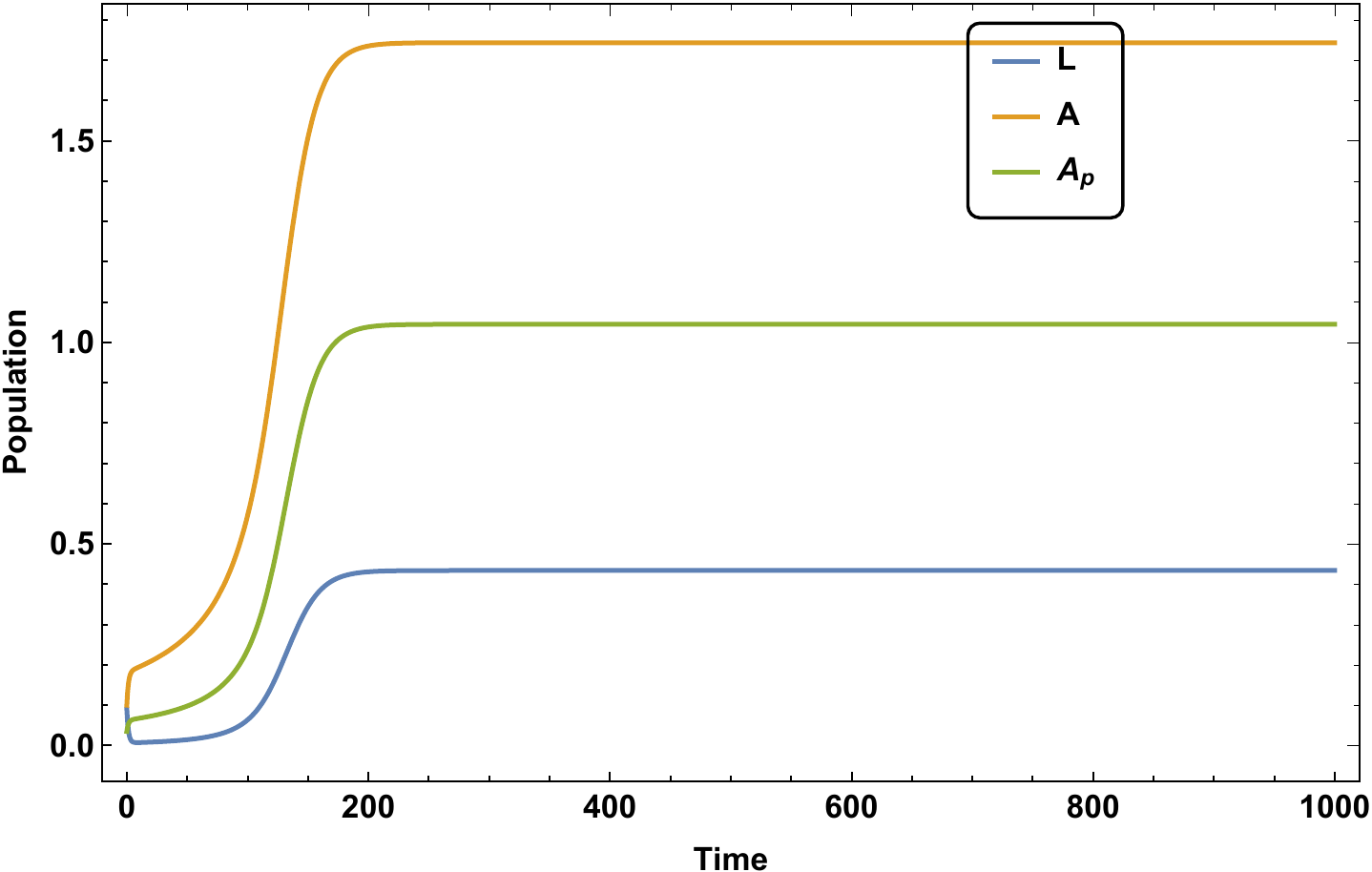}\label{time-1-2}}
\caption{{The time series of Model~\eqref{M1} when $\theta_c=2\theta_m=8$ with an initial value $(L(0),A(0),A_p(0))=(0.09,0.1,0.035)$ and other parametric values being $b=0.1,\,d= 0.1,\,\alpha=0.3,\,\beta = 0.7,\,\gamma = 0.9,$ which is the same set of parameters values in Figure~\ref{Figbif-a}. Figure~\ref{time-1-1} is the case when $a=0.07$ where population goes extinct over time. Figure~\ref{time-1-2} is the case when $a=0.1$ where the colony has a locally asymptotically stable interior equilibrium $E_2=(0.434, 1.743, 1.045)$}.}\label{Figtimephase} 
\end{figure}

Figure \ref{Figtimephase} provides two examples of population dynamics of Model~\eqref{M1} to show the effects of $a$. The initial condition is  $(L(0),A(0),A_p(0))=(0.09,0.1,0.035)$ and other parameters values are the same as in Figure~\ref{Figbif-a}.  According to Theorem~\ref{th-Survival-new}, Model~\eqref{M1} has global stability at $E_0=(0,0,0)$ (i.e., colony collapses) if $a=0.07<\min\big\{a^*,\frac{bd}{\alpha\gamma\theta_m}\big\}$ (see Figure~\ref{time-1-1}) and Model~\eqref{M1} has  two positive interior equilibria with $E_2=(0.434, 1.743, 1.045)$ being a locally asymptotically stable interior equilibrium if $a=0.1>a^*$ (see Figure~\ref{time-1-2}). It indicates that larvae $L$, worker ants $A$, and the ants collecting proteinaceous material $A_p$ can coexist with proper initial conditions if $a$ is in the intermediate range.

\textbf{Symmetrical case $\theta_c\leq2\theta_m$ (i.e., $\theta_0=2\theta_m-\theta_c>0$):} Figure~\ref{Figbif-Aa} provides an example of bifurcation diagram on division of labor invested on larvae ($a$) of Model~\eqref{M1} by  taking same parameter values in Figure~\ref{Figbif-a} except that $\theta_c=7.8$. Figure~\ref{Figbif-Aa} shows that Model~\eqref{M1} undergoes a bifurcation as the portion of the division of labor invested on larvae $a$ decreasing past {$\tilde{a}^*=\frac{4bd^2\beta^2(1-\alpha)^2}{\alpha^4\gamma^2+4d\alpha\beta^2\gamma(1-\alpha)(\alpha-(1-\alpha)(2\theta_m-\theta_c))}$}. Notice that the symmetrical case has  $\theta_0=2\theta_m-\theta_c>0$. Thus, the comparisons between Figure~\ref{Figbif-Aa}  and  Figure~\ref{Figbif-a} can provide insights on the effect of $a$ and $\theta_c$ (or $\theta_0$ because we set $\theta_m=4$ and $\theta_0=2\theta_m-\theta_c$): (1) The smaller value of $\theta_c$, the larger critical threshold  ${a}^*$; (2) The smaller value of $\theta_c$, the smaller population $A$.  The dynamical outcomes of symmetrical cases are similar to the general case shown in Figure~\ref{figgeneral} and Figure~\ref{fig-ApAc-a}.

\begin{figure}[!ht]
\centering
{\includegraphics[width=70mm]{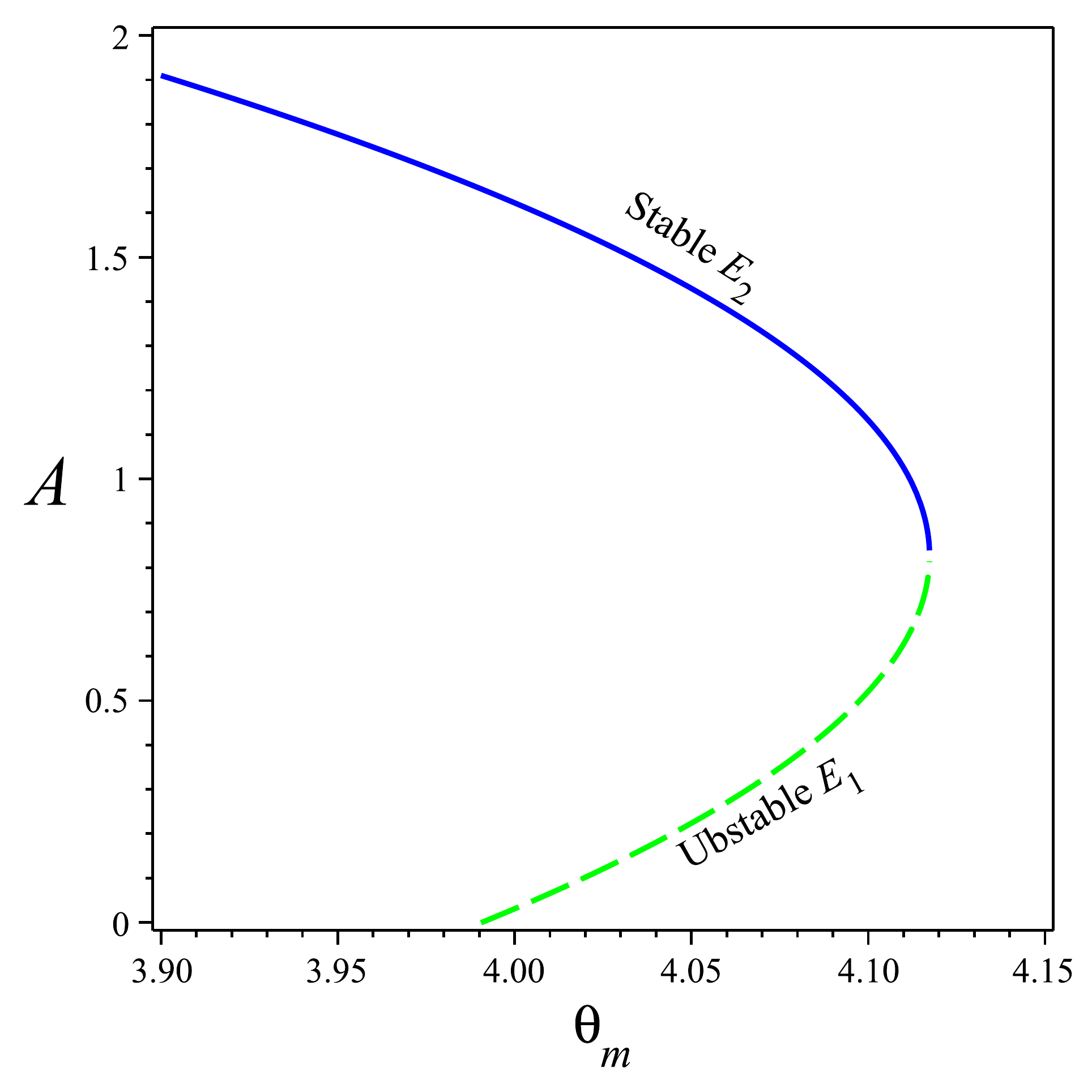}}
\caption{Bifurcation diagram for Model~\eqref{M1} on the optimal nutrient ratio $\theta_m$. An unstable interior equilibrium (green dotted) bifurcates $\theta_m^*=\frac{1}{2}\big(\theta_c+\frac{\alpha}{1-\alpha}+\frac{\alpha^3\gamma}{4d\beta^2(1-\alpha)^2}-\frac{bd}{a\alpha\gamma}\big)$.  The solid line indicates that the equilibrium is stable while the dotted line indicates that the equilibrium is unstable. Parameters values: $a = 0.15,\,b=0.1,\,d= 0.1,\,\alpha=0.3,\,\beta = 0.7,\,\gamma = 0.9,\,\,\theta_c =7.8$.}\label{Figbif-two}
\end{figure}

To understand the effects of the optimal nutrient ratio $\theta_m$ (or $\theta_0=2\theta_m-\theta_c=2\theta_m-7.8$), we perform a bifurcation diagram on $\theta_m$ shown in Figure~\ref{Figbif-two} by setting
  $${a = 0.15,\,b=0.1,\,d= 0.1,\,\alpha=0.3,\,\beta = 0.7,\,\gamma = 0.9,\,\,\theta_c =7.8.}$$
Notice that $\theta_0=2\theta_m-\theta_c\geq 0$, thus the value of $\theta_m$ in Figure~\ref{Figbif-two} starts with $\theta_m=1/2\theta_c=3.9$. Figure~\ref{Figbif-two} shows that Model \eqref{M1} exhibits reversed backward bifurcation on $\theta_m$ (or $\theta_0$):  (1) small values of the optimal nutrient ratio $\theta_m$ can insecure the persistence of the colony; (2) intermediate values of $\theta_m$ can go through saddle node bifurcation; and (3) large values of $\theta_m$ can lead to colony collapse.
 
\section{Conclusion}\label{S-Dis}

Variation in nutrient consumption among individuals is considered a conserved mechanism regulating castes and division of labor in social insects colonies. In eusocial insects, foragers, who perform food collection tasks, need to satisfy their own nutrient requirements in addition to those of the non-foraging workers, as well as the larvae and queen(s), which have significantly higher protein needs~\cite{HolldoblerWilson2009}. In this paper, we propose and study a nonlinear differential equations system to explore how nutritional status may regulate population dynamics and foraging task allocation of social insect colonies by applying adaptive modeling framework. Our model assumes that foragers adjust their preferences in favor of food sources containing limiting nutrients to maintain colony growth and reproduction~\cite{pohl2016colony,dussutour2008carbohydrate,clark2011behavioral}.

Our proposed model consists of a population of larvae $L$, foragers collecting carbohydrate $A_c$, and foragers collecting protein $A_p$. We assume that the survival rate of larvae is determined by the available nutrition in the colony, which is reflected through the ratio of workers collecting protein to those collecting carbohydrates ${S_{L_{\max}}}\left(\frac{A_p}{A_c}\right)$. Our formulation of ${S_{L_{\max}}}\left(\frac{A_p}{A_c}\right)$ is based on biological studies (see ~\cite{pohl2016colony,dussutour2008carbohydrate}) and embeds with an adaptive modeling approach adopted from~\cite{kang2011mathematical,Kang2014JMAM,Kang2015NRM}. Our theoretical results and bifurcation analysis conclude that  our proposed model exhibits backward bifurcations that generate bistability (see examples in Figure \ref{Figtimephase}). The bistability of the colony implies that initial conditions are important for colony survival under certain ranges of life history parameters. More specifically, the dynamical features  and the related biological implications of Model~\eqref{M1} can be summarized as follows:

\begin{enumerate}
\item The nutrition status measured by $\frac{A_p}{A_c}$ is an increasing function of the total population of workers $A$, or vice versa. This result may stem from the assumption that larvae (or brood in general) have higher nutritional needs to ensure survival. The biological implications of this are that higher nutritional status $\frac{A_p}{A_c}$, can lead to a better survival rate of larvae, thus the colony can grow with larger worker population $A$.
 \item The survival probability of brood is an increasing function of the following important life history parameters of colony:
 \begin{enumerate}
 \item The division of labor invested on brood measured by $a$ can have huge impacts on dynamical outcomes of the colony. From Lemma \ref{invariant}, Theorems~\ref{th-existence-general} and~\ref{th-Survival-general}, Model \eqref{M1} exhibits a backward bifurcation (shown in Figure~\ref{figgeneral}) as $a$ decreasing past its critical point 
 $$\hat{a}^*=\frac{4bd^2\beta^2(1-\alpha_1)^2}{\alpha_1^4\gamma^2+4d\alpha_1\beta^2\gamma(1-\alpha_1)(\alpha_1-(1-\alpha_1)\theta_0)}$$ which is an increasing function of the maturation rate $\beta$, the minimal nutrient ratio $\theta_0$; and a decreasing function of queen(s) laying egg rate $\gamma$. The larger the value of $a$, the more likely it is that the colony survives and grows. 
\item Effects of nutrient thresholds $\theta_0, \theta_m, \theta_c$:  Figure~\ref{figgeneral}, ~\ref{fig-ApAc-a} and~\ref{Figbif-two} suggest that Model 
\eqref{M1} exhibits reversed backward bifurcation on $\theta_m$ (or $\theta_0$): (1) small values of the optimal nutrient ratio $\theta_m$ can insecure the persistence of the colony; (2) intermediate values of $\theta_m$ can go through saddle node bifurcation that leads to bistability; and (3) large values of $\theta_m$ can lead to colony collapsing. 

In addition, the larger value of  the optimal nutrient ratio $\theta_m$ (or $\theta_0$) leads to (1)  larger critical threshold  $\hat{a}^*$; (2) smaller population $A$; and (3) smaller nutrient status, i.e., smaller value of  $\frac{A_p}{A_c}$.

\item Effects of the brood survival rate $\alpha_1$: Figure \ref{fig-LAAp-theta0} shows that Model \eqref{M1} exhibits reversed backward bifurcation on $\theta_0$. Figure \ref{fig-A-a-alpha} suggests that the larger value of $\alpha_1$, the larger the population of workers $A$ will be, which increases the probability of colony survival.
 
\end{enumerate}
\end{enumerate}

Our proposed model and study provide new insights into the strategies used by social insects (such as harvesting ants) facing nutritional challenges, and our results deepen our understanding of their nutritional ecology. Task allocation has been studied in social insects, that is, how colonies change the allocation of tasks in response to changing colony needs. One of future directions would be extending our current model to include more tasks such as brood care, foraging, and study how different tasks are related to colony needs including nutritional requirement. An another future direction is to extend our current model to include an additional level such as food resource of the colony. For example, leaf-cutter ants collect leaves as food resource to cultivate fungi and harvest the fruits of fungi as their food. The nutrient requirement of the leaf-cutter ants colony has two levels: one is the needs of the colony itself such as brood and the other one is the needs of the fungi. It would be interesting to explore how nutrient needs of the colony and fungus garden affect the foraging behavior of leaf-cutter ants during its ontology.  

\section{Proofs}\label{proofs}

\subsection*{Proof of Lemma \ref{invariant}}

\begin{proof}
\textcolor{black}{For any $L,\,A\in\mathbb{R}_+^2$, from Model~\eqref{M1} we obtain $L'|_{L(0)=0}\geq 0$ and $A'|_{A(0)=0}\geq 0$ for all $t\geq0$. Since $A=A_p+A_c$, then $A_p'|_{A_p(0)=0}\geq 0$ for all $t\geq0$. Moreover, if $L(0)=0,\,A(0)=0$ and $A_p(0)=0$, then $(L(t),A(t),A_p(t))=(0,0,0)$ for all $t\geq0$. If $L(0)>0,\,A(0)>0$ and $A_p(0)>0$, then by continuity arguments, it is impossible for either $L(t)$ or $A(t)$ or $A_p(t)$ to drop below $0$. Hence, for any $L(0)\geq0,\,A(0)\geq0$ and $A_p(0)\geq0$, we obtain $L(t)\geq0,\,A(t)\geq0$ and $A_p(t)\geq0$ for all $t\geq0$.}

\textcolor{black}{Now assume $L(0)\geq0,\,A(0)\geq0$ and $A_p(0)\geq0$, then since the function of $S_L(\frac{A_p}{A_c})$ exists maximum when $\frac{A_p}{A_c}\in\big(\max\{0,\theta_0\},\theta_c\big)$ and according to the expression of $L'$, we have
\begin{equation*}
L'=\alpha_1\gamma\left(\frac{A_p}{A_c}-\theta_0\right)\frac{aA^2}{b+aA^2}-\beta L\leq\alpha_1\gamma(\theta_m-\theta_0)-\beta L
\end{equation*}
for all $t\geq0$ when $\theta_m>\theta_0$. Thus, a standard comparison theorem shows that $\limsup_{t\rightarrow\infty} L(t)\leq \frac{\alpha_1\gamma(\theta_m-\theta_0)}{\beta}$. This indicates that for any $\epsilon>0$, there exists $T$ large enough, such that 
$$L(t)\leq \frac{\alpha_1\gamma(\theta_m-\theta_0)}{\beta}+\epsilon \quad \mbox{for all}\quad t>T.$$ Therefore, from the expression of $A'$, we have
\begin{equation*}
A'=\beta L-dA^2\leq \beta\left(\frac{\alpha_1\gamma(\theta_m-\theta_0)}{\beta}+\epsilon\right)-dA^2 \quad\mbox{for all}\quad t>T.
\end{equation*}
Since $\epsilon$ can be arbitrarily small, thus $\limsup_{t\rightarrow\infty} A(t)\leq\frac{\sqrt{d\alpha_1\gamma(\theta_m-\theta_0)}}{d}$. Thus, we have shown that the Model \eqref{M1} is positively invariant and bounded in $\mathbb{R}_+^2$. More specifically, the compact set $\Big[0,\frac{\alpha_1\gamma(\theta_m-\theta_0)}{\beta}\Big]\times\Big[0,\frac{\sqrt{d\alpha_1\gamma(\theta_m-\theta_0)}}{d}\Big]$ attracts all points in $\mathbb{R}_+^2$. Due to $A=A_p+A_c$ and the boundedness of $A(t)$, hence we can obtain $A_p$ is bounded for all $t\geq0$.}

Moreover, if $L(0)>0,\,A(0)>0$ and $A_p(0)>0$, then we have follows:
\begin{equation*}
\begin{array}{l}
L'=\alpha_1\gamma\left(\frac{A_p}{A_c}-\theta_0\right)\frac{aA^2}{b+aA^2}-\beta L\geq-\beta L\Rightarrow L(t)\geq L(0)e^{-\beta t}>0,\\[8pt]
A'=\beta L-dA^2\geq-dA^2 \Rightarrow A(t)\geq \frac{A(0)}{1+dt}>0,\\[8pt]
A_p'=\alpha_1\beta L+\alpha_1 A_cL-{dA_p A}\geq-{dA_p A} \Rightarrow A_p(t)\geq A_p(0)e^{-d\int_0^tA(s)\mathrm{d}s}>0.
\end{array}
\end{equation*}
Therefore, if $L(0)>0,\,A(0)>0$ and $A_p(0)>0$, then $L(t)>0,\,A(t)>0$ and $A_p(t)>0$ for all $t>0$.
\end{proof}

\subsection*{Proof of Theorem \ref{th-existence-general}}

\begin{proof}
{It is easy to see that $E_0=(0,0,0)$ is always an equilibrium of Model \eqref{M1}. The nullclines of \eqref{M1} can be found as
\begin{equation*}
\begin{array}{l}
L'=0\Longrightarrow  \alpha_1\gamma\big(\frac{A_p}{A_c}-\theta_0\big)\frac{aA^2}{b+aA^2}-\blue{\beta L}=0,\\[8pt]
A'=0\Longrightarrow L=\frac{d}{\beta}A^2,\\[8pt]
A_p'=0\Longrightarrow \alpha_1\beta L+\alpha_1 A_cL-dA_p A=0\Longrightarrow L=\frac{dA_pA}{\alpha_1\beta+\alpha_1 A_c}.
\end{array}
\end{equation*}
By solving $\frac{dA_pA}{\alpha_1\beta+\alpha_1 A_c}=\frac{d}{\beta}A^2$ for $A_p$, we have $A_p=\frac{\alpha_1\beta A+\alpha_1 A^2}{\beta+\alpha_1 A}$ and substitute it to $L'=0$, which results in the following equation:
\begin{equation}\label{eq-A-general}
ad\beta(1-\alpha_1)A^2-a\alpha_1^2\gamma A+\beta[(1-\alpha_1)(bd+a\alpha_1\gamma\theta_0)-a\alpha_1^2\gamma]=0.
\end{equation}
The roots of \eqref{eq-A-general} are given by
\[A_1=\frac{a\alpha_1^2\gamma-\sqrt{{\Delta}}}{2a\beta d(1-\alpha_1)},\quad  
A_2=\frac{a\alpha_1^2\gamma+\sqrt{{\Delta}}}{2a\beta d(1-\alpha_1)},\]
where ${\Delta}=a(a\alpha_1^4\gamma^2-4d\beta^2[(1-\alpha_1)^2(bd+a\alpha_1\gamma\theta_0)-a\alpha_1^2\gamma(1-\alpha_1)])$.\\}

Thus, we have the following three cases:

{\noindent Let $\theta_0^*=\frac{\alpha_1^3\gamma}{4d\beta^2(1-\alpha_1)^2}+\frac{\alpha_1}{1-\alpha_1}-\frac{bd}{a\alpha_1\gamma}$.
\begin{enumerate}
\item If $\theta_0>\theta_0^*$ and $a>\frac{4bd^2\beta^2(1-\alpha_1)^2}{\alpha_1^2\gamma(\alpha_1^2\gamma+4d\beta^2(1-\alpha_1))}$, then there is only one trivial equilibrium: $E_0=(0,0,0)$ and no other positive interior equilibrium.
\item If $\theta_0=\theta_0^*$ and $a>\frac{4bd^2\beta^2(1-\alpha_1)^2}{\alpha_1^2\gamma(\alpha_1^2\gamma+4d\beta^2(1-\alpha_1))}$, then Model \eqref{M1} has two positive equilibria which collapse into one equilibrium $E_*$ as
  \[({L}_*,A_*,{A_p}_*)=\left(\frac{d}{\beta}A_*^2,\frac{\alpha_1^2\gamma}{2\beta d(1-\alpha_1)},\frac{\alpha_1\beta A_*+\alpha_1 A_*^2}{\beta+\alpha_1A_*}\right).\] Or if $0<\theta_0<\frac{\alpha_1}{1-\alpha_1}-\frac{bd}{a\alpha_1\gamma}$ and $a>\frac{bd(1-\alpha_1)}{\alpha_1^2\gamma}$, then Model \eqref{M1} has only one positive equilibrium 
\[({L}_2,A_2,A_{p2})=\left(\frac{d}{\beta}A_2^2, A_2,\frac{\alpha_1\beta A_2+\alpha_1A_2^2}{\beta+\alpha_1 A_2}\right).\]
\item If $\max\big\{0,\frac{\alpha_1}{1-\alpha_1}-\frac{bd}{a\alpha_1\gamma}\big\}<\theta_0<\theta_0^*$ and $a>\frac{bd(1-\alpha_1)}{\alpha_1^2\gamma}$, then Model \eqref{M1} has two positive equilibria in the following form:
\[
  ({L}_1,A_1,A_{p1})=\left(\frac{d}{\beta}A_1^2, A_1,\frac{\alpha_1\beta A_1+\alpha_1A_1^2}{\beta+\alpha_1A_1}\right)\quad\text{and}\quad
  ({L}_2,A_2,A_{p2})=\left(\frac{d}{\beta}A_2^2, A_2,\frac{\alpha_1\beta A_2+\alpha_1A_2^2}{\beta+\alpha_1A_2}\right).\]
\end{enumerate}}
\end{proof}

\subsection*{Proof of Theorem \ref{extinction-E0-general}}
\begin{proof}
From Proposition~\ref{Prop-1}, we know that for some initial condition taken in $\mathbb{R}_+^3$ and if $a<\frac{bd}{\alpha_1\gamma(\theta_m-\theta_0)}$, the trajectory of Model~\eqref{M1} is converging to the origin $E_0=(0,0,0)$.  And according to Theorem~\ref{th-existence-general}, if $0<a<\hat{a}^*$ and $\theta_0<\frac{\alpha_1^3\gamma+4d\alpha_1\beta^2(1-\alpha_1)}{4d\beta^2(1-\alpha_1)^2}$, then there is only one trivial equilibrium $E_0=(0,0,0)$ and no other positive equilibrium. Therefore, we can conclude that Model~\eqref{M1} has global stability at $(0,0,0)$ when $a<\min\big\{\hat{a}^*,\frac{bd}{\alpha_1\gamma(\theta_m-\theta_0)}\big\}$ and $\theta_0<\frac{\alpha_1^3\gamma+4d\alpha_1\beta^2(1-\alpha_1)}{4d\beta^2(1-\alpha_1)^2}$.
\end{proof}

\subsection*{Proof of Theorem \ref{th-Survival-general}}

\begin{proof}
The local stability of equilibria is determined by computing the eigenvalues of the Jacobian matrix about each equilibrium.

{Let $E^*=(L^*,A^*,A_p^*)$ be an arbitrary positive equilibrium of Model~\eqref{M1}. The Jacobian matrix at this equilibrium is
\begin{equation}\label{Jacobian-general}
\begin{array}{l}
J|_{E^*}=\left(
                      \begin{array}{ccc}
                        -\beta & J_{12} & J_{13} \\
                        \beta & -2dA^* & 0 \\
                        J_{31} & J_{32} & J_{33}\\
                      \end{array}
                    \right),
\end{array}
\end{equation}
where
\begin{equation*}
\begin{array}{l}
J_{12}=\frac{a\alpha_1\gamma A^*[A_p^*(bA^*-aA^{*3}-2bA_p^*)-2b(A^*-A_p^*)^2\theta_0]}{(A^*-A_p^*)^2(b+aA^{*2})^2},\quad\quad  
J_{13}=\frac{a\alpha_1\gamma A^{*3}}{(A^*-A_p^*)^2(b+aA^{*2})}>0,\\[6pt]
J_{31}=\alpha_1(\beta+A^*-A_p^*)>0, \quad\quad 
J_{32}=\alpha_1 L^*-dA_p^*,\quad\quad
J_{33}= -(\alpha_1L^*+d A^*)<0.
\end{array}
\end{equation*}
Then we have the characteristic equation of $J|_{E^*}$ is
\begin{equation}\label{characteristic-general}
\hat{f}(\lambda)=\lambda^3+C_1\lambda^2+C_2\lambda+C_3=0,
\end{equation}
where
\begin{equation*}
\begin{array}{l}
C_1=\beta+\alpha L^*+3d A^*>0,\\[4pt]
C_2=J_{11}J_{33}+J_{11}J_{22}+J_{22}J_{33}-J_{21}J_{12}-J_{31}J_{13}\\[4pt]\hspace{0.4cm}
=2d\beta A^*+(\beta+2dA^*)(\alpha_1L^*+dA^*)-\frac{a\alpha_1^2\gamma A^{*3}(\beta+A^*-A_p^*)}{(A^*-A_p^*)^2(b+aA^{*2})}\\[4pt]\hspace{0.5cm}
-\frac{a\alpha_1\beta\gamma A^*\big[A_p^*(bA^*-aA^{*3}-2bA_p^*)-2b(A^*-A_P^*)^2\theta_0\big]}{(A^*-A_p^*)^2(b+aA^{*2})^2},\\[8pt]
C_3=-\textrm{det}(J|_{E_i^*})=J_{11}J_{22}J_{33}+J_{21}J_{32}J_{13}-J_{21}J_{12}J_{33}-J_{31}J_{22}J_{13}\\[4pt]\hspace{0.4cm}
=-2d\beta A^*(\alpha_1L^*+dA^*)+\frac{a\alpha_1\gamma A^{*3}[\beta(\alpha_1L^*-dA_p^*+2d\alpha_1A^*)+2d\alpha_1A^*(A^*-A_p^*)]}{(A^*-A_p^*)^2(b+aA^{*2})}\\[4pt]\hspace{0.5cm}
+\frac{a\alpha_1\beta\gamma A^*(\alpha_1L^*+dA^*)\big[A_p^*(bA^*-aA^{*3}-2bA_p^*-2b(A^*-A_p^*)^2\theta_0)\big]}{(A^*-A_p^*)^2(b+aA^{*2})^2}.
\end{array}
\end{equation*}
This indicates the following two cases:
\begin{enumerate}
\item If Model~\eqref{M1} has a unique interior equilibrium 
$E_2=(L_2,A_2,A_{p2})=\left(\frac{d}{\beta}A_2^2, A_2,\frac{\alpha_1\beta A_2+\alpha_1A_2^2}{\beta+\alpha_1A_2}\right)$, then under the conditions $0<\theta_0<\frac{\alpha_1}{1-\alpha_1}-\frac{bd}{a\alpha_1\gamma}$ and $a>\frac{bd(1-\alpha_1)}{\alpha_1^2\gamma}$ and $C_1(E_2)C_2(E_2)>C_3(E_2)>0$, thus, by applying the Routh-Hurwitz criterion, we can obtain that the interior equilibrium $E_2$ of Model~\eqref{M1} is locally asymptotically stable.
\item If Model~\eqref{M1} has two interior equilibria $E_i= (L_i,A_i,A_{pi})=\left(\frac{d}{\beta}A_i^2, A_i,\frac{\alpha_1\beta A_i+\alpha_1A_i^2}{\beta+\alpha_1A_i}\right),\,i=1,2$ where $E_1<E_2$, then under the conditions $\max\big\{0,\frac{\alpha_1}{1-\alpha_1}-\frac{bd}{a\alpha_1\gamma}\big\}<\theta_0<\theta_0^*$, $a>\frac{4bd^2\beta^2(1-\alpha_1)^2}{\alpha_1^2\gamma(\alpha_1^2\gamma+4d\beta^2(1-\alpha_1))}$, and $C_1(E_1)C_2(E_1)-C_3(E_1)<0$ but $C_1(E_2)C_2(E_2)>C_3(E_2)>0$, we can obtain that the interior equilibrium $E_2$ is locally asymptotically stable while $E_1$ is unstable.
\end{enumerate}}
\end{proof}

\subsection*{Proof of Theorem \ref{th-Survival-new}}

\begin{proof}
For any $L,\,A,\,A_p\in\mathbb{R}_+^3$, note that 
\begin{equation*}
\begin{array}{l}
L'|_{L=0}=\alpha\gamma\frac{A_p}{A-A_p}\frac{aA^2}{b+aA^2}\geq 0,\\[4pt]
A'|_{A=0}=\beta L\geq 0,\\[4pt]
A_p'|_{A_p=0}=\alpha\beta L+\alpha AL\geq 0,
\end{array}
\end{equation*}
thus according to Theorem A.4 (p. 423) of~\cite{Thieme2003book}, we can conclude that the model \eqref{M1} is positive invariant in $\mathbb{R}_+^3$. Now we can proceed {to} show the boundedness of the system. First, assume $L(0)\geq0,\,A(0)\geq0$ and $A_p(0)\geq0$, then since the function of $S_L(\frac{A_p}{A_c})$ exists maximum when $0<\frac{A_p}{A_c}\leq\theta_m$ and according to the expression of $L'$, we have the following inequalities due to the property of positive invariance:
\[L'=\alpha\gamma\frac{A_p}{A_c}\frac{aA^2}{b+aA^2}-\beta L\leq \alpha\gamma\theta_m-\beta L \]
which implies that
\[\limsup_{t\rightarrow\infty}L(t)\leq\frac{\alpha\gamma\theta_m}{\beta}. \]
This suggests that there exists $\epsilon>0$ such that the following inequalities hold as time $t$ is large enough,
\[A'=\beta L-dA^2\leq\beta\Big(\frac{\alpha\gamma\theta_m}{\beta}+\epsilon\Big)-dA^2  \]
which indicates that 
\[\limsup_{t\rightarrow\infty}A(t)\leq\sqrt{\frac{\alpha\gamma\theta_m}{d}}. \]
Then, we also have the following inequalities hold as time $t$ is large enough,
\[A_p'=\alpha\beta L+\alpha(A-A_p)L-dAA_p\leq\alpha\Big(\frac{\alpha\gamma\theta_m}{\beta}+\epsilon\Big)\Big(\beta+\sqrt{\frac{\alpha\gamma\theta_m}{d}}+\epsilon\Big)-d\Big(\sqrt{\frac{\alpha\gamma\theta_m}{d}}+\epsilon\Big)A_p\]
which shows that 
\[ \limsup_{t\rightarrow\infty}A_p(t)\leq\alpha\Big(\sqrt{\frac{\alpha\gamma\theta_m}{d}}+\frac{\alpha\gamma\theta_m}{d\beta}\Big). \]
Therefore, every trajectory starting from $\mathbb{R}_+^3$ converges to the compact set $$\mathbb{C}=\big[0,\frac{\alpha\gamma\theta_m}{\beta}\big]\times\big[0,\sqrt{\frac{\alpha\gamma\theta_m}{d}}\big]\times\big[0,\alpha\big(\sqrt{\frac{\alpha\gamma\theta_m}{d}}+\frac{\alpha\gamma\theta_m}{d\beta}\big)\big].$$

Let $E^*=(L^*,A^*,A_p^*)$ be an interior equilibrium of Model~\eqref{M1}. Then its stability is determined by the eigenvalues $\lambda_i(E^*),\,i=1,2,3$ of its associated Jacobian matrix as follows:
\begin{equation*}\label{Jacobian-Ei}
\begin{array}{l}
J|_{E^*}=\left(
                      \begin{array}{ccc}
                        -\beta & -\frac{a\alpha\gamma A^*A_p^*(a{A^*}^3-bA^*+2bA_p^*)}{(A^*-A_p^*)^2(b+a{A^*}^2)^2} & \frac{a\alpha\gamma {A^*}^3}{(A^*-A_p^*)^2(b+a{A^*}^2)} \\
                        \beta & -2dA^* & 0 \\
                         \alpha(\beta+A^*-A_p^*)  & \alpha L^*-dA_p^* & -\alpha L^*-dA^*\\
                      \end{array}
                    \right),
\end{array}
\end{equation*}
since $\beta L^*=d{A^*}^2,\>\beta L^*=\alpha\gamma\frac{A_p^*}{A_c^*}\frac{a{A^*}^2}{b+a{A^*}^2}$, and $\frac{A^*}{A_p^*}=\frac{\beta}{\alpha(\beta+A_c^*)}$. Therefore, we have 
\begin{equation*}\label{Jacobian-Ei-2}
\begin{array}{l}
J|_{E^*}=\left(
                      \begin{array}{ccc}
                        -\beta & -\frac{d^2{A^*}^2(a{A^*}^2-b)}{a\alpha\gamma A_p^*}-\frac{2bd^2A^*}{a\alpha\gamma} & \frac{d{A^*}^3}{A_p^*A_c^*} \\
                        \beta & -2dA^* & 0 \\
                         \alpha(\beta+A^*-A_p^*)  & \alpha L^*-dA_p^* & -\alpha L^*-dA^*\\
                      \end{array}
                    \right),
\end{array}
\end{equation*}
and the characteristic equation of $J|_{E^*}$ is
\begin{equation*}\label{characteristic-new}
f(\lambda)=\lambda^3+c_1\lambda^2+c_2\lambda+c_3=0,
\end{equation*}
where
\begin{equation*} 
\begin{array}{l}
c_1=-\textrm{tr}(J|_{E^*})=-(\lambda_1(E^*)+\lambda_2(E^*)+\lambda_3(E^*))=\beta+3dA^*+\alpha L^*>0,\\[4pt]
c_2=(\beta+2dA^*)(\alpha L^*+dA^*)+\frac{\beta d^2A^*}{a\alpha\gamma}\big(\frac{A^*(a{A^*}^2-b)}{A_p^*}+2b\big)+2d\beta A^*-\frac{\beta d {A^*}^2}{A_c^*},\\[4pt]
c_3=-\textrm{det}(J|_{E^*})=-\lambda_1(E^*)\lambda_2(E^*)\lambda_3(E^*)\\[4pt]\hspace{0.4cm}
=\big[\frac{\beta d^2A^*}{a\alpha\gamma}(\frac{A^*(a{A^*}^2-b)}{A_p^*}+2b)+2d\beta A^*\big](\alpha L^*+dA^*)-\frac{\beta d{A^*}^3}{A_p^*A_c^*}(\alpha L^*-dA_p^*)-\frac{2\beta d^2{A^*}^3}{A_c^*}.
\end{array}
\end{equation*}
This indicates the following two cases:
\begin{enumerate}
\item If Model~\eqref{M1} has a unique interior equilibrium 
$E_2=(L_2,A_2,{A_p}_2)=\left(\frac{d}{\beta}A_2^2, A_2,\frac{\alpha\beta A_2+\alpha A_2^2}{\beta+\alpha A_2}\right)$, then under the conditions $a>\frac{bd(1-\alpha)}{\alpha^2\gamma}$, $a_1<a<a_2$ and 
\begin{equation*} 
\begin{array}{l} 
\max\big\{0,\frac{\beta}{\beta-\alpha A_2}\big\}<\frac{A_2}{{A_c}_2}<\min\left\{\frac{\alpha L_2+dA_2}{\alpha L_2+d{A_p}_2}\big[\frac{d(a{A_2}^2-b)}{a\alpha\gamma}+\frac{2{A_p}_2}{A_2}(\frac{bd}{a\alpha\gamma}+1)\big],M\right\},
\end{array}
\end{equation*}
where 
\begin{equation}\label{a12-M}
\begin{array}{l} 
a_1=\frac{b\big(\alpha^2\gamma+4d\beta^2(1-\alpha)-\alpha\sqrt{\gamma(\alpha^2\gamma+8d\beta^2(1-\alpha))}\big)}{2\alpha^2\beta^2\gamma},\\[4pt]\,a_2=\frac{b\big(\alpha^2\gamma+4d\beta^2(1-\alpha)+\alpha\sqrt{\gamma(\alpha^2\gamma+8d\beta^2(1-\alpha))}\big)}{2\alpha^2\beta^2\gamma},\\[4pt]
M=\frac{1}{\beta^2}(2\alpha d{A_2}^2+(\alpha+2d)\beta A_2+3\beta^2)+\frac{d}{a\alpha\gamma}\big(\frac{A_2}{{A_p}_2}(a{A_2}^2-b)+2b\big),
\end{array}
\end{equation}
we get $c_2>0,\,c_3>0$. And we can verify that $c_1c_2-c_3>0$. Thus, we can conclude that the interior equilibrium $E_2$ is locally stable by applying the Routh-Hurwitz criterion.
\item If Model~\eqref{M1} has two interior equilibria $E_i= (L_i,A_i,{A_p}_i)=\left(\frac{d}{\beta}A_i^2, A_i,\frac{\alpha\beta A_i+\alpha A_i^2}{\beta+\alpha A_i}\right),\,i=1,2$ where $E_1<E_2$, then under the conditions $a^*<a<\frac{bd(1-\alpha)}{\alpha^2\gamma}$, $a_1<a<a_2$, and 
\begin{equation*} 
\begin{array}{l} 
\frac{A_1}{{A_c}_1}>\frac{1}{\beta^2}(2\alpha d{A_1}^2+(\alpha+2d)\beta A_1+3\beta^2)-\frac{d}{a\alpha\gamma}\big(\frac{A_1}{{A_p}_1}(b-a{A_1}^2)-2b\big), \\[4pt]
\frac{A_2}{{A_c}_2}<\frac{1}{\beta^2}(2\alpha d{A_2}^2+(\alpha+2d)\beta A_2+3\beta^2)+\frac{d}{a\alpha\gamma}\big(\frac{A_2}{{A_p}_2}(a{A_2}^2-b)+2b\big),
\end{array}
\end{equation*}
we obtain $A_1<\sqrt{\frac{b}{a}}<A_2$ and $c_2(E_1)<0$ but $c_2(E_2)>0$. We also can verify that $c_1(E_2)c_2(E_2)-c_3(E_2)>0$. Therefore, the interior equilibrium $E_2$ is locally asymptotically stable while $E_1$ is unstable.
\end{enumerate}
\end{proof}

\section*{Acknowledgements}
This research of F.R. is partially supported by the National Science Foundation of China (Grant Nos. 11601226 \& 11426132 \& 71871115), Qing Lan Project of Jiangsu Province, the Natural Science Foundation of Jiangsu Province of China (Grant No. BK20140927), and the research funds from Nanjing Tech University and Jiangsu Government Scholarship for Overseas Studies. The work of Y.K. is also partially supported by NSF-DMS (1313312\& 1716802); NSF-IOS/DMS (1558127), DARPA (ASC-SIM II), and The James S. McDonnell Foundation 21st Century Science Initiative in Studying Complex Systems Scholar Award (UHC Scholar Award 220020472).

\bibliographystyle{spmpsci}
\bibliography{antscitation}

\end{document}